\documentclass[aps, prx, reprint, preprintnumbers, superscriptaddress, amssymb]{revtex4-1}

\usepackage{graphicx}
\usepackage{dcolumn}
\usepackage{bm}

\begin{document}

\preprint{Ver. 9}¡¡

\title{Fermi Surface with Dirac Fermions in CaFeAsF Determined via Quantum Oscillation Measurements }


\author{Taichi Terashima}
\author{Hishiro T. Hirose}
\affiliation{National Institute for Materials Science, Tsukuba, Ibaraki 305-0003, Japan}
\author{David Graf}
\affiliation{National High Magnetic Field Laboratory, Florida State University, Tallahassee, FL 32310, USA}
\author{Yonghui Ma}
\author{Gang Mu}
\author{Tao Hu}
\affiliation{State Key Laboratory of Functional Materials for Informatics, Shanghai Institute of Microsystem and Information Technology, Chinese Academy of Sciences, Shanghai 200050, China}
\affiliation{CAS Center for Excellence in Superconducting Electronics (CENSE), Shanghai 200050, China}
\author{Katsuhiro Suzuki}
\affiliation{Research Organization of Science and Technology, Ritsumeikan University, Kusatsu, Shiga 525-8577, Japan}
\author{Shinya Uji}
\affiliation{National Institute for Materials Science, Tsukuba, Ibaraki 305-0003, Japan}
\author{Hiroaki Ikeda}
\affiliation{Department of Physics, Ritsumeikan University, Kusatsu, Shiga 525-8577, Japan}


\date{\today}
\begin{abstract}
Despite the fact that 1111-type iron arsenides hold the record transition temperature of iron-based superconductors, their electronic structures have not been studied much because of the lack of high-quality single crystals.
In this study, we compehensively determine the Fermi surface in the antiferromagnetic state of CaFeAsF, a 1111 iron-arsenide parent compound, by performing quantum oscillation measurements and band-structure calculations.
The determined Fermi surface consists of a symmetry-related pair of Dirac electron cylinders and a normal hole cylinder.
From analyses of quantum-oscillation phases, we demonstrate that the electron cylinders carry a nontrivial Berry phase $\pi$.
The carrier density is of the order of 10$^{-3}$ per Fe.
This unusual metallic state with the extremely small carrier density is a consequence of the previously discussed topological feature of the band structure which prevents the antiferromagnetic gap from being a full gap.
We also report a nearly linear-in-$B$ magnetoresistance and an anomalous resistivity increase above about 30 T for $B \parallel c$, the latter of which is likely related to the quantum limit of the electron orbit.
Intriguingly, the electrical resistivity exhibits a nonmetallic temperature dependence in the paramagnetic tetragonal phase ($T >$ 118 K), which may suggest an incoherent state.
Our study provides a detailed knowledge of the Fermi surface in the antiferromagnetic state of 1111 parent compounds and moreover opens up a new possibility to explore Dirac-fermion physics in those compounds.
\end{abstract}


\maketitle



\newcommand{\ud}{\mathrm{d}}
\def\degree{\kern-.2em\r{}\kern-.3em}

\section{introduction}
Since the discovery of superconductivity at $T_c$ = 26 K in LaFeAs(O$_{1-x}$F$_x$) by Kamihara \textit{et al.} \cite{Kamihara08JACS}, iron-based superconductors have been studied extensively.
By replacing La with smaller rare-earth elements, $T_c$ of up to 56 K was achieved in the same 1111-type structure immediately after the discovery \cite{Kito08JPSJ, Ren08CPL, Yang08SST, Wang08EPL}.
There are now various structure types: the 122- (e.g. BaFe$_2$As$_2$ \cite{Rotter08PRL, Sasmal08PRL}), 111- (e.g. LiFeAs \cite{Tapp08PRB, Pitcher08ChemCom, Wang08SSC}), 11-type (e.g. FeSe \cite{Hsu08PNAS}) structures and so on.
Nonetheless, the above $T_c$ record has not yet been surpassed by other structure types as far as bulk materials are concerned.
It is therefore crucial to unravel the normal-state electronic structure of the 1111 iron arsenides from which the highest $T_c$ among the iron-based superconductors emerges.
Unfortunately, however, such studies suffer from difficulty in obtaining high-quality single crystals of the 1111-type iron arsenides.

There is a variant of the 1111 iron arsenides, i.e., CaFeAsF (or SrFeAsF) \cite{Matsuishi08JACS, Tegel08EPL, Han08PRB} that has the same ZrCuSiAs-type structure as LaFeAsO, but the LaO layers in LaFeAsO are replaced by the CaF layers (Fig. \ref{struct}).
The electronic band structure near the Fermi level $E_F$ is predicted to be similar to that of LaFeAsO \cite{Shein08JETPLett}.
CaFeAsF exhibits a structural and an antiferromagnetic transition as LaFeAsO does \cite{Tegel08EPL, Xiao09PRB_CaFeAsF}.
The antiferromagnetic structure is the same stripe-type one as LaFeAsO \cite{Xiao09PRB_CaFeAsF}.
By partial substitution of Fe by Co, Ca(Fe, Co)AsF exhibits superconductivity with a maximum $T_c$ of 22 K \cite{Matsuishi08JACS}.
Further, $T_c$ of 56 K has been observed for rare-earth doped compounds \cite{Wu09JPCM, Cheng09EPL}.
Following initial reports of single-crystal growth of (Ca, Na)FeAsF \cite{Shlyk14SST} and CaFeAsF$_{1-x}$ \cite{Tao14SciChina}, Ma \textit{et al}. have succeeded in growing large high-quality single crystals of undoped CaFeAsF by self-flux method \cite{Ma15SST}. 

In this article, we report resistivity and magnetic-torque measurements down to 0.03 K (0.4 K) up to 18 T (45 T) performed on high-quality single crystals of CaFeAsF.
The temperature dependence of the resistivity is nonmetallic down to $T_s$ = 118 K and exhibits two clearly separated anomalies at $T_s$ and $T_N$ = 107 K, which are attributed to a structural and an antiferromagnetic transition, respectively.
We observe clear quantum oscillations both in the resistivity and magnetic torque in the antiferromagnetic ground state.
Two frequencies $\alpha$ and $\beta$ are resolved, and their field-angle dependences indicate that they are from highly two-dimensional Fermi-surface cylinders.
From detailed analyses of the oscillation phases, we find a nontrivial Berry phase $\pi$ associated with the $\alpha$ oscillation, demonstrating that the $\alpha$ cylinder is a Dirac-fermion cylinder.
Our band-structure calculations indicate that the Fermi surface consists of a hole cylinder at the $\Gamma$ point of the Brillouin zone and a symmetry-related pair of electron cylinders originating from Dirac lines located away from $\Gamma$.
Thus the $\alpha$ and $\beta$ cylinders are attributed to the electron and hole cylinders, respectively.
The carrier density is extremely small, of the order of 10$^{-3}$ per Fe.
Additionally, we report a nearly linear-in-$B$ magnetoresistivity and an anomalous resistivity increase above about 30 T for $B \parallel c$.
Among other things, our work opens up a new possibility to explore Dirac-fermion physics in a branch of the family of the iron-based superconductors with a simple band structure.

\section{Methods}
High-quality single crystals of CaFeAsF were prepared by a CaAs self-flux method in Shanghai \cite{Ma15SST}.
The crystals were platelike with an $ab$-plane area of up to $\sim$1 mm$^2$ and a thickness of $\sim$50 $\mu$m or less. 

\begin{figure}
\includegraphics[width=8.6cm]{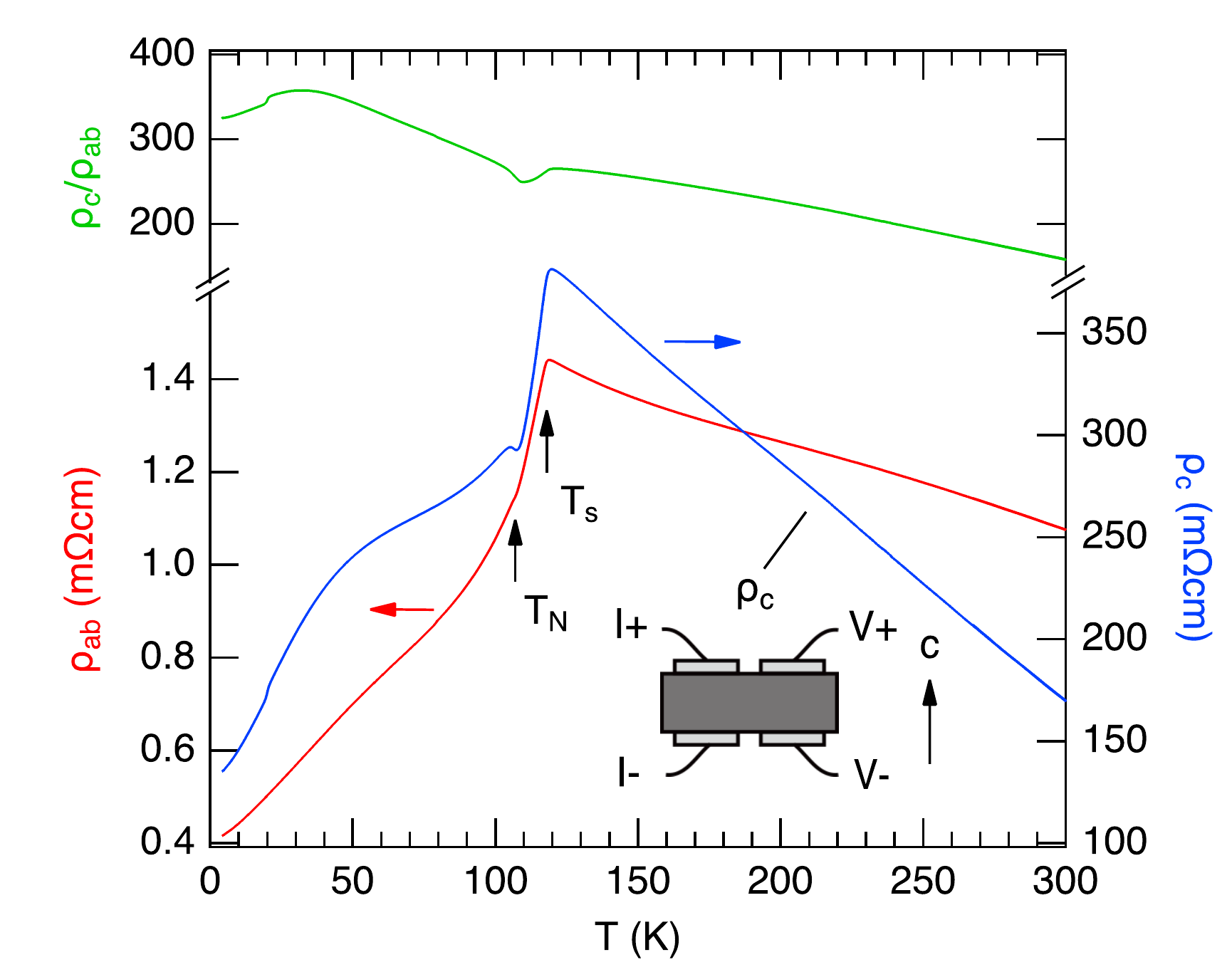}
\caption{\label{RvsT}(Color online).  Temperature dependence of resistivity in CaFeAsF.  The $c$-axis and $ab$-plane resistivities were measured using two different samples.  The inset explains the contact configuration for the $c$-axis measurements.  The upper panel shows the anisotropy $\rho_c/\rho_{ab}$.}   
\end{figure}

A 20-T superconducting magnet with a dilution refrigerator at the National Institute for Materials Science and the 45-T hybrid magnet with a $^3$He refrigerator at the National High Magnetic Field Laboratory were used.
In either magnet, samples can be rotated in situ, and the field angle $\theta$ is measured from the $c$ axis. 
Electrical resistivity $\rho$ measurements were performed with a four-contact method both for current $I \parallel ab$ and $I \parallel c$ (see Fig.~\ref{RvsT} for the contact configuration for $I \parallel c$).
The contacts were spot-welded or made with conducting silver paste.
A total of six samples were prepared: two $ab$-plane and one $c$-axis samples were used for SdH measurements, while the remaining one $ab$-plane and two $c$-axis samples were used only for zero-field temperature-dependence measurements.
Microcantilever torque $\tau$ measurements were performed on another two crystals.
Consistent quantum-oscillation results were obtained for the three SdH and two torque crystals.

\begin{figure}
\includegraphics[width=4cm]{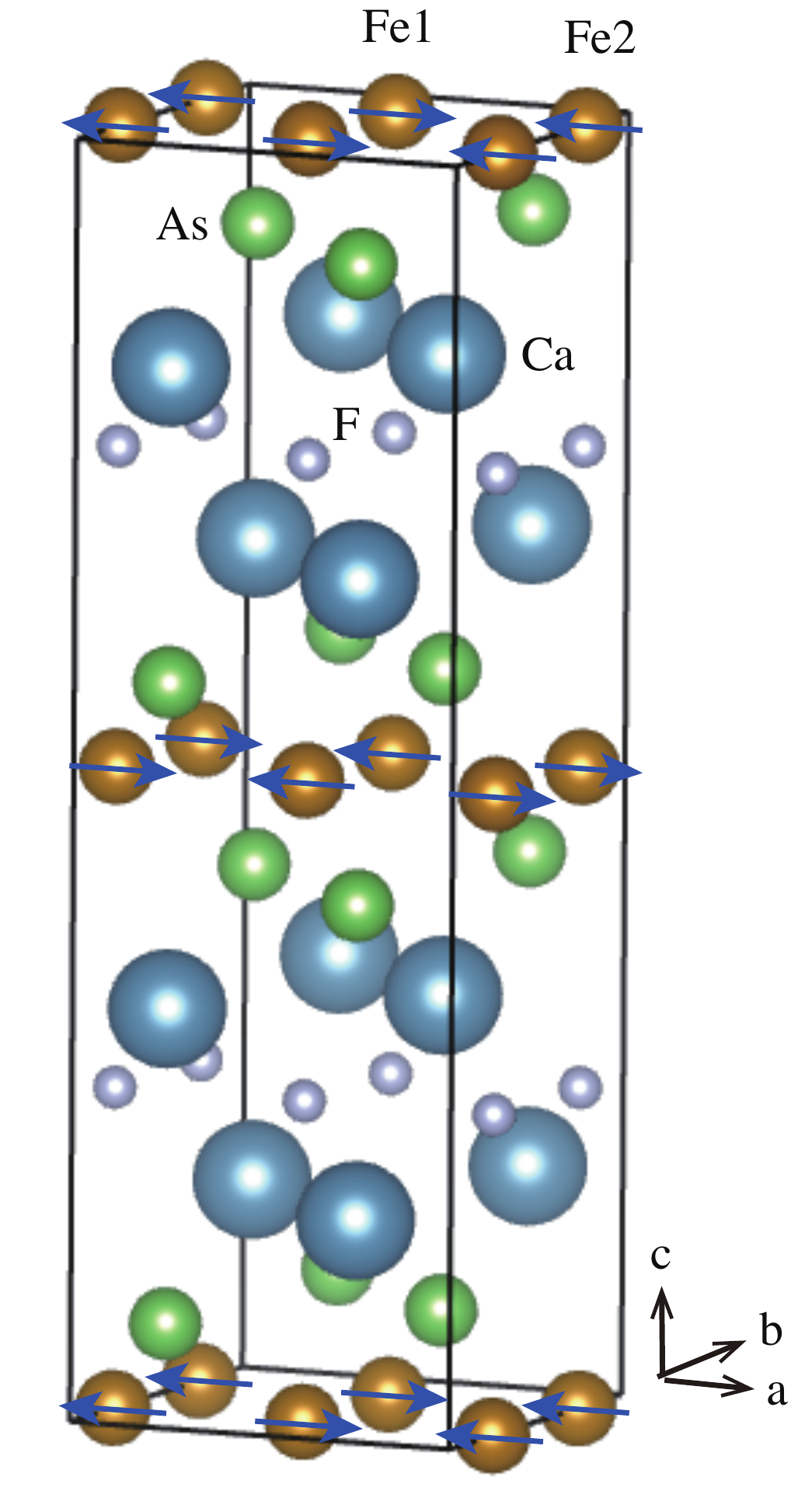}
\caption{\label{struct}  Crystal and magnetic structure of CaFeAsF.  The $a$ , $b$, and $c$ directions of the orthorhombic Cmma cell are indicated.   The black frame indicates the magnetic cell with the doubled $c$ dimension, and the arrows indicate the Fe moment direction.  Fe1 and Fe2 label magnetic sublattices used to calculate partial density of states in Fig. \ref{band}.}   
\end{figure}

Fully relativistic electronic band-structure calculations with spin-orbit coupling were performed by using the WIEN2k code with the PBE-GGA potential \cite{WIEN2K, Perdew96PRL}.
We used the experimental lattice parameters of the orthorhombic crystal structure (space group Cmma) at $T$ = 2 K: $a$ = 5.514594 \AA, $b$ = 5.476802 \AA, $c$ = 8.576845 \AA, $z_{As}$ = 0.66329 \cite{Xiao09PRB_CaFeAsF}\footnote{Y. Xiao (private communication).}.
The antiferromagnetic ordering with $\mathbf{Q}$ = (1,0, 1/2) (w.r.t Cmma) \cite{Xiao09PRB_CaFeAsF} was incorporated by using space group Ibam (\#72) with a doubled $c$ dimension (Fig. \ref{struct}).
In the self-consistent calculations, the muffin-tin radii $R_{\rm{MT}}$ of Ca, Fe, As and F atom were set to 2.15, 2.31, 2.19, and 2.26 in units of the Bohr radius, respectively.
The maximum modulus of reciprocal vectors was chosen such that $R_{\rm MT}K_{\rm max}=7.0$.
The corresponding Brillouin zone was sampled using a $10 \times 10 \times 10$ $k$-mesh for the self-consistent calculations, and $40 \times 40 \times 40$ $k$-mesh for calculations in the density of states.

\section{Temperature dependence of resistivity}
Figure \ref{RvsT} shows the in-plane and $c$-axis resistivities as a function of temperature $T$.
The resistivities increase initially as $T$ is lowered from room temperature as reported in \cite{Ma15SST}.
These measurements were made on the very same high-quality single crystals that exhibit quantum oscillations at low temperatures.
This indicates that the observed non-metallic behavior is intrinsic nature of CaFeAsF.

The resistivities decrease sharply below $T_s$ = 118 K and exhibit a kink at $T_N$ = 107 K as reported in \cite{Ma15SST}.
These anomalies at $T_s$ and $T_N$ most likely correspond to a structural and an antiferromagnetic phase transition, respectively, as observed in a neutron powder-diffraction experiment by Xiao \textit{et al} \cite{Xiao09PRB_CaFeAsF}. 
The $c$-axis resistivity of this sample exhibits a sudden faint drop at $T$ = 20 K, which is not observed in the other samples and hence is probably due to a crack.
The resistivity anisotropy is large and basically increases with decreasing temperature.

The room-temperature in-plane and $c$-axis resistivities, averaged out over the three in-plane and three $c$-axis samples, are 1.9(8) m$\Omega$cm and 2.3(6)$\times10^2$ m$\Omega$cm, respectively. 
The large errors can be ascribed to errors in the sample dimensions, irregular sample shapes and inhomogeneous current distribution.
The large anisotropy $\rho_c/\rho_{ab}$ of the order of 10$^2$ is broadly consistent with the large upper-critical-field anisotropy of 9 observed for CaFe$_{0.882}$Co$_{0.118}$AsF \cite{Ma17SST} as the two anisotropies are approximated by $m^*_c/m^*_{ab}$ and $\sqrt{m^*_c/m^*_{ab}}$, respectively.

\section{quantum oscillations}
\subsection{Theoretical background}
Quantum oscillation of the resistivity $\Delta\rho$ can be described by \cite{Richards73PRB, Shoenberg84} 
\begin{equation}
\frac{\Delta\rho}{\rho_0} = -C\sqrt{B}R_TR_DR_s \cos\left[2\pi\left(\frac{F}{B}-\frac{1}{2}\right)+\phi_D+\phi_B\right],
\end{equation}
where $\rho_0$ is the background resistivity, $C$ a positive coefficient, and we have neglected harmonics.
We have tacitly assumed that $\sigma = \rho/(\rho^2+\rho_H^2) \approx \rho^{-1}$ since the Hall resistivity $\rho_H$ is much smaller than $\rho$ \cite{Cheng09EPL}.
The frequency $F$ is related to a Fermi-surface cross sectional area $A$ as $F = (\hbar/2\pi e)A$.
The temperature and Dingle reduction factors are given by $R_T= X/\sinh X$ and $R_D=\exp(-X_D)$, where $X_{(D)}=K\mu^* T_{(D)}/B$, $\mu^*=m^*/m_e$, and the coefficient $K$ is 14.69 T/K.
The Dingle temperature $T_D$ is inversely proportional to the carrier scattering time $\tau$: $T_D=\hbar / 2\pi k_B\tau$.
The spin reduction factor $R_s$ describes the interference of oscillations from up- and down-spin electrons and is given by $R_s=\cos (\pi g \mu^*/2)$, where $g$ is the spin $g$ factor.
$\phi_D$ is 0 for a two-dimensional (2D) Fermi-surface (FS) cylinder while it is + or $-\pi/4$ when the oscillation is from a minimum or maximum cross section of a three-dimensional (3D) FS pocket.
$\phi_B$ is the Berry phase, which is 0 for normal electrons but $\pi$ for Dirac fermions \cite{Mikitik99PRL, Mikitik98JETP}.
The sign of $R_s$ is essential to determine the Berry phase in systems where electronic bands are doubly degenerate at each $k$ due to spin degrees of freedom as is the case with the present one \cite{*[{}] [{.  Even if a system under consideration is antiferromagnetically ordered, the double degeneracy remains as long as the product of the inversion (and a translation if necessary) and the time reversal is a symmetry of the system.}] Cvetkovic13PRB}.
If $R_s > 0$ and $\phi_D$ is neglected, resistivity maxima occur for normal fermions when $F/B$ takes integer values, while minima for Dirac fermions.
If $R_s < 0$, the opposite is true.

\subsection{Frequency and effective mass}
\begin{figure}
\includegraphics[width=8.6cm]{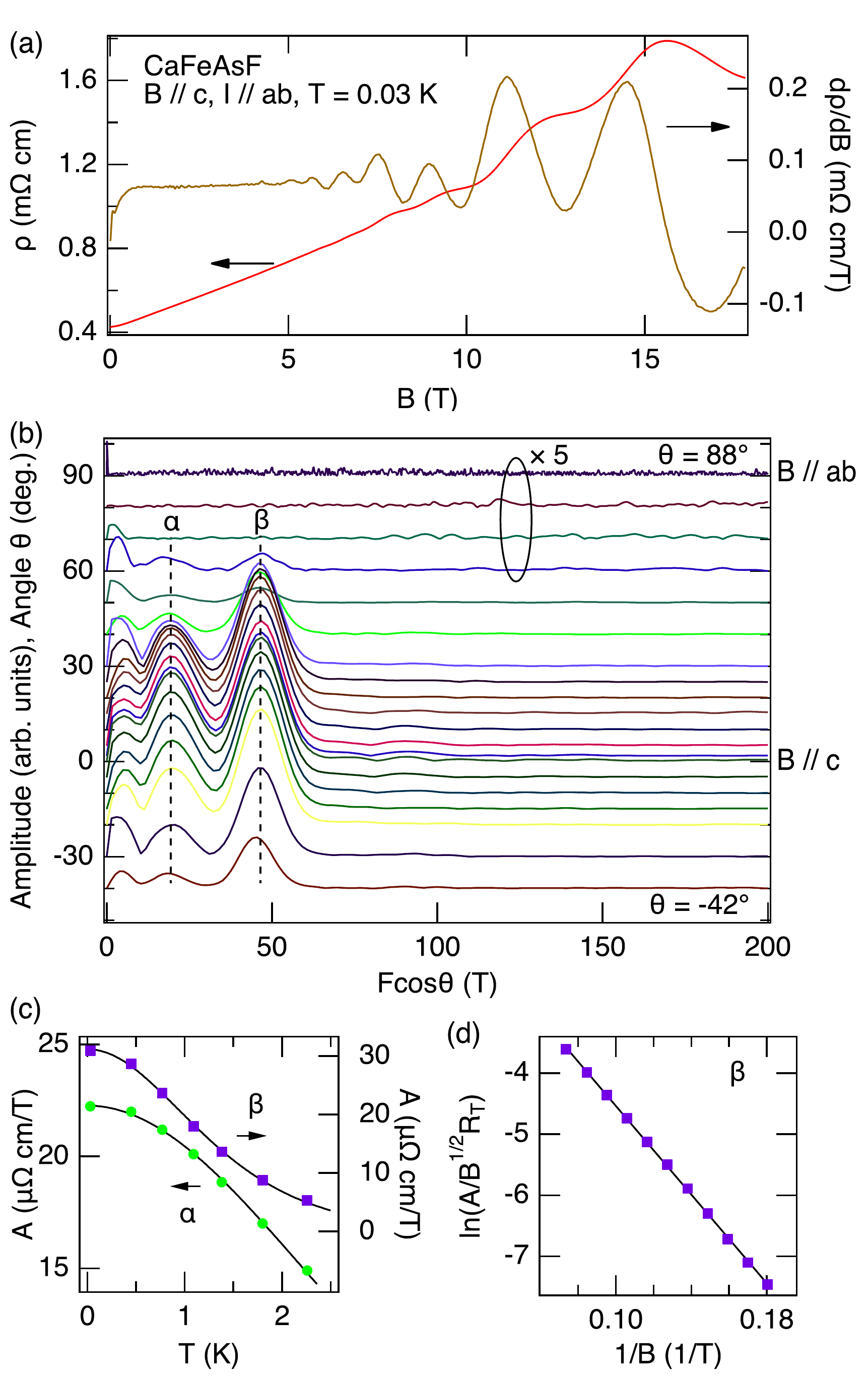}
\caption{\label{sig20T}  (a) In-plane resistivity in CaFeAsF and its magnetic-field derivative as a function of $B$ applied parallel to the $c$ axis.  (b) Fourier transforms of d$\rho$/d$B$ vs $1/B$ over the field range 5--17.8 T for various field directions $\theta$.  Note that the horizontal axis is $F\cos\theta$.  The spectra are vertically shifted so that the baseline of a spectrum for an angle $\theta$ is placed at $\theta$ of the vertical axis.  (c) Temperature dependence of the $\alpha$ and $\beta$ oscillation amplitudes for $B \parallel c$.  The curves are fits to the temperature reduction factor $R_T$.  (d) Dingle plot for the $\beta$ frequency for $B\parallel c$.  The vertical axis is logarithm of a reduced amplitude.  The line is a linear fit.}   
\end{figure}

Figure \ref{sig20T}(a) shows the in-plane resistivity $\rho$ and field derivative d$\rho$/d$B$ as a function of $B$ up to 17.8 T applied along the $c$ axis.
The field derivative shows an initial rise up to $B \sim$0.6 T, above which it stays approximately constant except oscillations.
Namely, the magnetoresistance is nearly linear in $B$.

Quantum oscillations are discernible above $B \sim$ 5 T.
Figure \ref{sig20T}(b) shows Fourier transforms of oscillations as a function of $F\cos\theta$ for various field directions $\theta$.
Two frequencies $\alpha$ and $\beta$ are resolved: $F_{\alpha}$ = 19.6 T and $F_{\beta}$ = 46.7 T for $B \parallel c$, which correspond to 0.14 and 0.34\% of the cross section of the antiferromagnetic Brillouin zone.
For both frequencies, $F\cos\theta$ stays constant for the investigated $\theta$ range, indicating that they are from highly 2D FS cylinders.
The effective masses associated with the $\alpha$ and $\beta$ orbits for $B \parallel c$ are estimated from the temperature dependences of the oscillation amplitudes to be 0.383(4) and 0.92(1), respectively, in units of the free electron mass $m_e$ [Fig. \ref{sig20T}(c)].
Figure \ref{sig20T}(d) shows the field dependence of the $\beta$ oscillation amplitude in the form of a Dingle plot.
The amplitudes were estimated from the extremums of the pure $\beta$ oscillation obtained below in Sec. \ref{phase}.
From the linear fit to the data, $T_D$ = 2.7 K, corresponding to a carrier mean free path of 21 nm ( $l = \tau v_F$, see Table I for $v_F$).

\begin{figure}
\includegraphics[width=8.6cm]{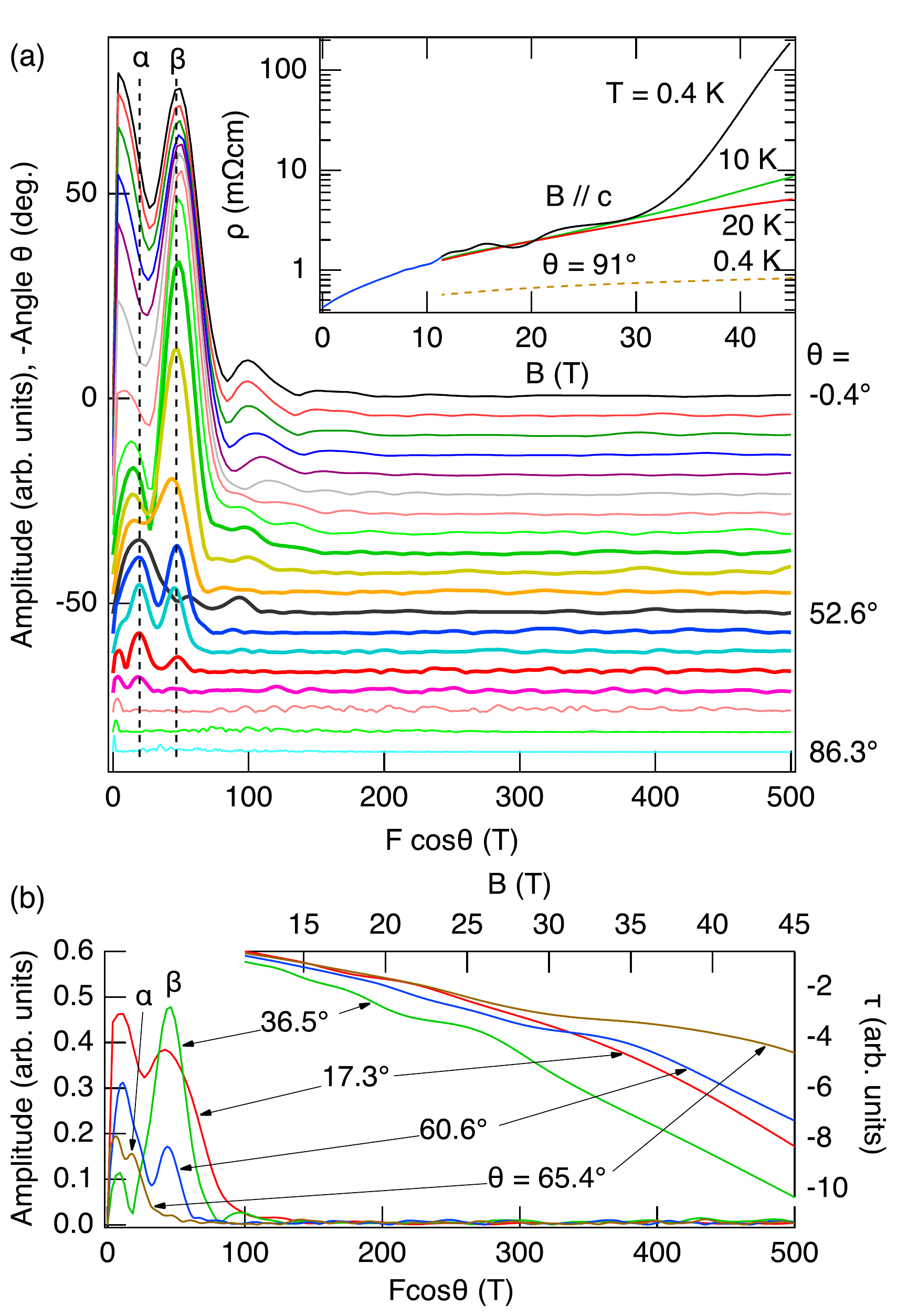}
\caption{\label{Hybrid}  (a, inset) In-plane resistivity in CaFeAsF as a function of $B$ up to 45 T for $B \parallel c$ and nearly $B \parallel ab$ ($\theta$ = 91$^{\circ}$).  (a, main) Fourier transforms of d$\ln\rho$/d$B$ vs $1/B$ over the field range 11.4--44.7 T for various field directions $\theta$.  Note that the horizontal axis is $F\cos\theta$.  The spectra are vertically shifted as in Fig. \ref{sig20T}(b).  (b) Magnetic torque $\tau$ in CaFeAsF (upper right) and corresponding Fourier transforms of d$\tau$/dB vs $1/B$ plotted against $F\cos\theta$.}   
\end{figure}

In order to search for higher frequencies, i.e., larger FS pockets, we extended measurements up to $B$ = 45 T by using the hybrid magnet.
The inset of Fig. \ref{Hybrid}(a) shows semilogarithmic plots of in-plane resistivity vs field.
For $B \parallel c$, the resistivity at $T$ = 0.4 K shows a remarkable increase above $B\sim$ 30 T.
The increase is suppressed as the temperature is raised, and almost disappears at $T$ = 20 K.
For nearly $B \parallel ab$ ($\theta$ = 91$^{\circ}$), no such an increase is observed.
A similar behavior is observed for the $c$-axis resistivity (data not shown).
Although the origin of the anomalous resistivity increase is not clear at present, it may be related to the quantum limit (i.e., the limit where only the lowest Landau level is occupied) of the $\alpha$ orbit ($F_{\alpha}$ = 19.6 T).
Apart from the anomalous increase, the undulation of the $T$ = 0.4 K curve for $B \parallel c$ is dominated by the $\beta$ oscillation: the maxima at $B$ $\sim$ 12, 16, and 23 T are ascribable to it, and no faster oscillation is seen.

The main panel of Fig. \ref{Hybrid}(a) shows Fourier transform analysis of resistivity data measured in the hybrid magnet.
We have analyzed the logarithmic derivative $\ud\ln\rho/\ud B$ instead of d$\rho$/d$B$ to suppress huge background due to the anomalous resistivity increase near $B \parallel c$.
The spectra show peaks corresponding to $\alpha$, $\beta$, and their harmonics: peaks around 100 T are ascribable to 2$\beta$ and/or 6$\alpha$. 
No higher fundamental frequency than $\beta$ exists.

We also measured magnetic torque [Fig.\ref{Hybrid}(b)].
The undulation of the torque curves is again basically explicable with the $\beta$ oscillation, and no faster oscillation is seen.
The Fourier spectra confirm that there is no frequency higher than $\beta$.
To summarize, our experimental results indicate that the Fermi surface is composed of only two types of extremely small cylinders $\alpha$ and $\beta$.

\subsection{\label{phase} Oscillation phase}
\begin{figure}
\includegraphics[width=8.6cm]{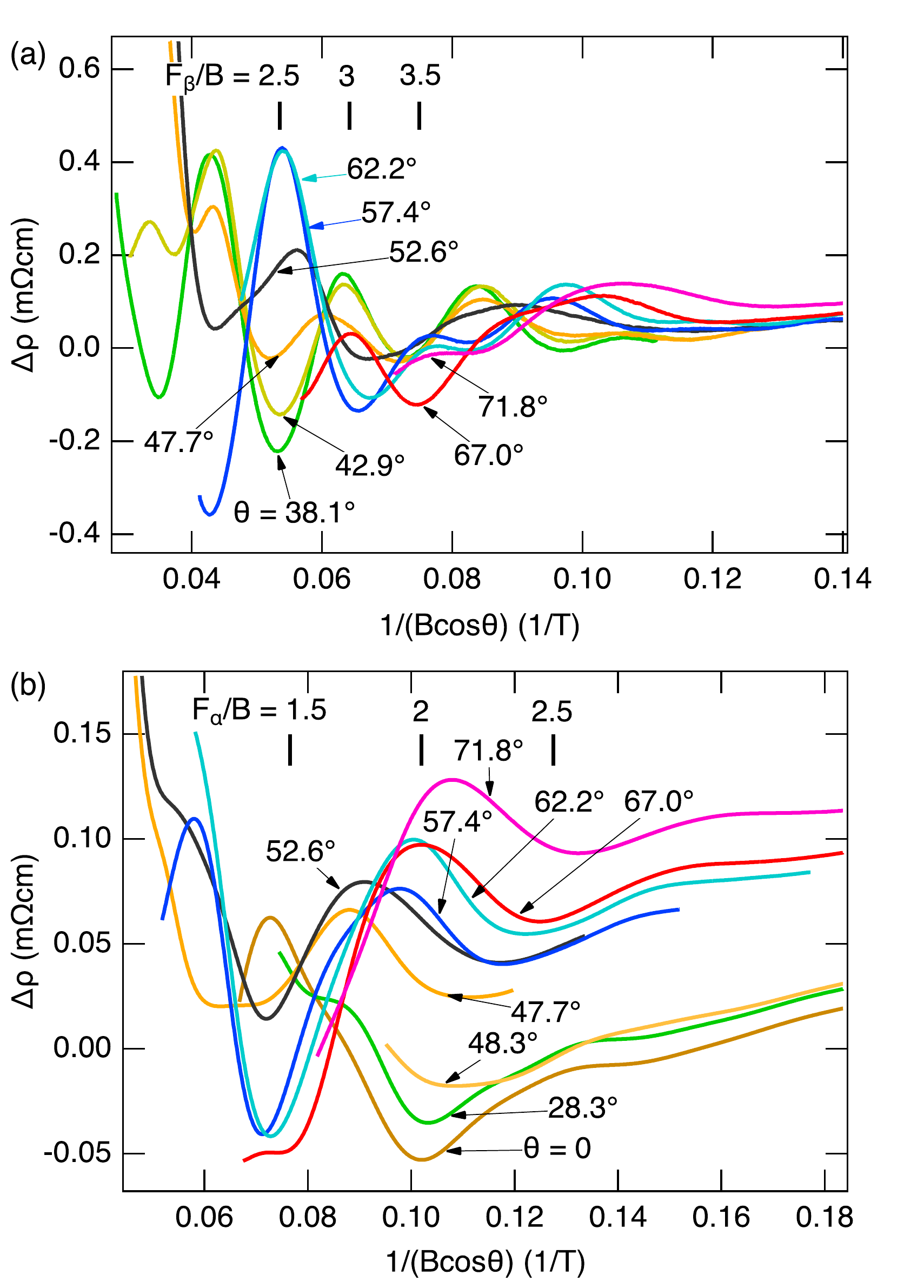}
\caption{\label{invBcos}  (a) Oscillatory part of the resistivity as a function of 1/($B\cos\theta$) for field directions near $\theta$ = 52.6$^{\circ}$.  All the curves are from the hybrid measurements.  (b) Oscillatory part of the resistivity as a function of 1/($B\cos\theta$) after box smoothing.  The box (window) size is the same as one oscillation period of the $\beta$ frequency, and hence the $\beta$ oscillation is effectively removed.  The curves for $\theta$ = 0, 28.3, and 48.3$^{\circ}$ are from the 20-T measurements, while the others from the hybrid measurements.}   
\end{figure}

We now consider the spin reduction factors $R_s$ and oscillation phases by analyzing the waveform of the observed oscillations.
To avoid ambiguity due to background subtraction, we subtract the same linear background $a+bB\cos\theta$ with $a$ = 260 $\mu\Omega$ cm and $b$ = 89.9 $\mu\Omega$ cm/T from the raw $\rho(B)$ data.
The parameters were determined by fitting a line to the $\rho(B)$ curve of Fig. \ref{sig20T}(a) in the field range 5--17.8 T.

For a 2D FS cylinder, the effective mass varies as 1/cos$\theta$ as the field direction $\theta$ is varied.
$R_s$ varies accordingly and may cross zero to change sign.
For such a ``spin-zero'' ($R_s = 0$), $g\mu^* = 1+2n$ where $n$ is an integer.

We begin with $R_s$ for the $\beta$ oscillation.
The Fourier transforms in Fig. \ref{Hybrid}(a) show that the $\beta$ amplitude almost vanishes at $\theta$ = 52.6$^{\circ}$ but revives at 57.4$^{\circ}$, suggesting a spin-zero near 52.6$^{\circ}$.
Figure \ref{invBcos}(a) shows the oscillatory part of the resistivity for field directions around $\theta$ = 52.6$^{\circ}$.
The curves at $\theta$ = 38.1, 42.9, and 47.7$^{\circ}$ show a minimum near $F_{\beta}/B = 2.5$, while those at 52.6, 57.4, and 62.2$^{\circ}$ a maximum: that is, the spin-zero exists between $\theta$ = 47.7 and 52.6$^{\circ}$.
Since the fundamental $\beta$ oscillation is weak at these angles ($\theta$ = 47.7 and 52.6$^{\circ}$) because of $R_s \approx 0$, the waveform is strongly influenced by the harmonics and the $\alpha$ oscillation, which explains the fact that the peak positions at these angles slightly differ from the other angles.
Similarly, looking at $F_{\beta}/B = 3.5$, we notice the second spin-zero between $\theta$ = 62.2 and 67.0$^{\circ}$ and the third between 67.0 and 71.8$^{\circ}$.

\begin{figure}
\includegraphics[width=8.6cm]{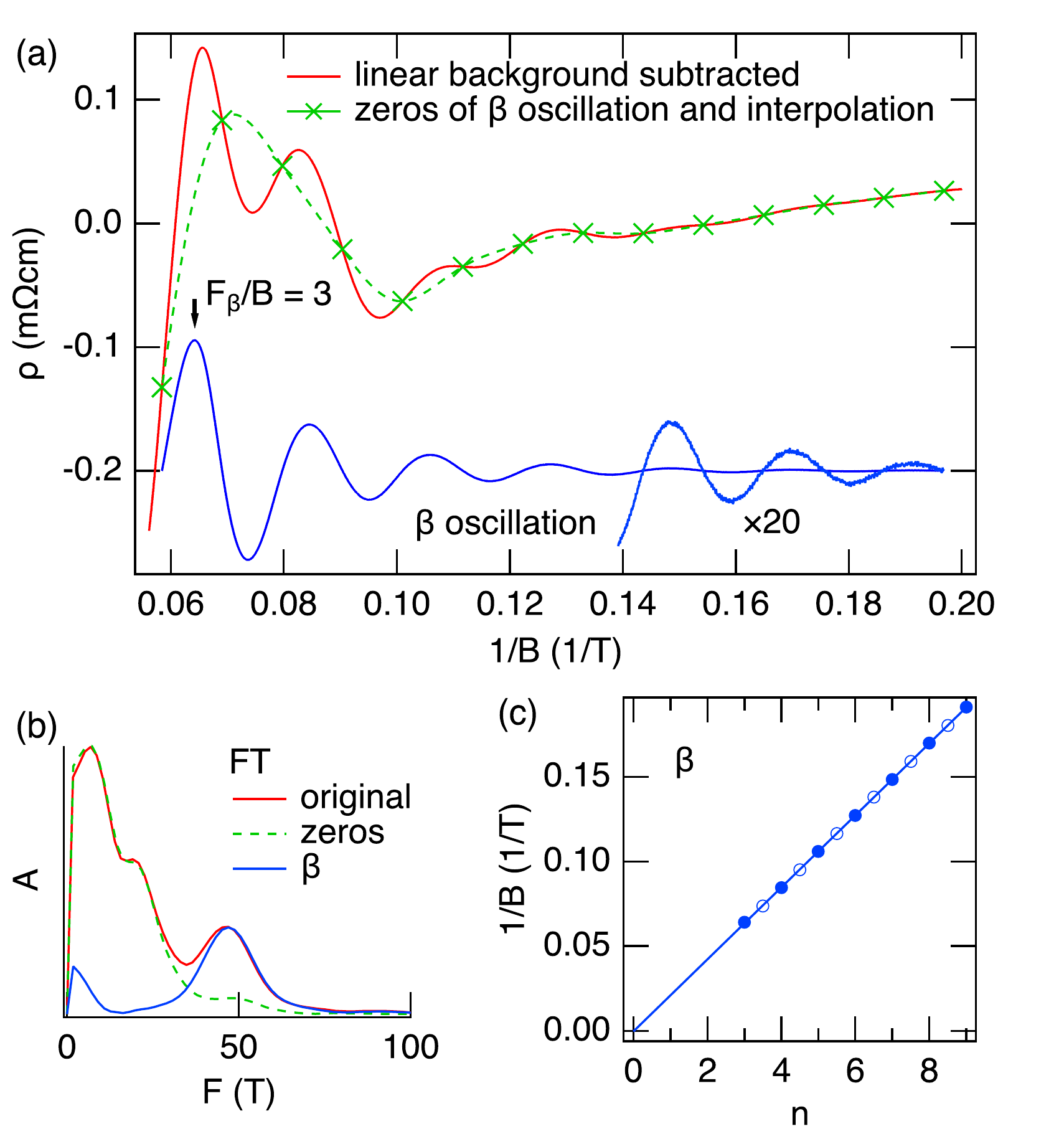}
\caption{\label{Decomp}  (a) Oscillatory part of the resistivity for $B \parallel c$ as a function of $1/B$ (solid curve).  The original data is the same as that of Fig. \ref{sig20T}(a).  The crosses are zeros of the $\beta$ oscillation determined from d$^2\rho$/d$B^2$, and the connecting curve (dashed curve) is obtained by cubic spline interpolation.  The $\beta$ oscillation (lower curve, vertically shifted) is obtained by subtracting the interpolated curve from the original.  (b) Fourier transforms of the original, interpolation, and $\beta$-oscillation curves.  (c) Landau-index plot for the $\beta$ frequency.  Integer and half integer indices are assigned to resistivity maxima (solid circles) and minima (open circles), respectively.}   
\end{figure}

Let us assume that the first spin-zero is at $\theta = 50^{\circ}$, where $\mu^*$ = 1.43 and hence $g = 0.699(1+2n)$.
If $n$ = 2 ($g$ = 3.49), the second and third spin-zeros occur at $\theta$ = 62.6 and 69.1$^{\circ}$, which are consistent with the experimental observations.
No other choice of $n$ is consistent with the experiment: $n < 2$ cannot explain the second spin-zero while $n > 2$ produces extra spin-zeros.  
Since $\mu^*$ = 0.92 at $\theta$ = 0, $\pi g \mu^*/2 = 1.61\pi$ and hence $R_s > 0$.
The conclusion $R_s > 0$ at $\theta$ = 0 is confirmed by a more general analysis which shows that the experimental observations can be explained with a $g$ value between 3.30 and 3.55, corresponding to the first spin-zero between 49.2 and 52.6$^{\circ}$.
Since $\Delta\rho$ shows a maxima at $F_{\beta}/B = 3$ at low angles where $R_s > 0$ [Figs. \ref{invBcos}(a) and \ref{Decomp}(a)], the $\beta$ carriers are normal fermions ($\phi_B$ = 0).

We now consider $\phi_D$ for the $\beta$ orbit.
Figure \ref{Decomp} shows the oscillatory part of the resistivity for $B \parallel c$ as a function of $1/B$.
We decompose this data into the $\beta$ oscillation and the rest, the latter of which contains the $\alpha$ oscillation, as follows:
We determine the positions of the zeros of the $\beta$ oscillation (crosses in the figure) by interpolating the maximum and minimum positions of d$^2\rho$/d$B^2$, which are dominated by the $\beta$ oscillation.
We apply cubic-spline interpolation to them, subtract the interpolated curve (dashed curve) from the data, and obtain the pure $\beta$ oscillation (lower curve).
The Fourier transforms in Fig. \ref{Decomp}(b) confirms the success of the decomposition.
Figure \ref{Decomp}(c) shows a Landau-index plot based on the resistivity maxima and minima of the thus obtained $\beta$ oscillation.
Integer and half integer indices are assigned to resistivity maxima and minima, respectively.
A linear fit (solid line) gives a slope of 0.0213 T$^{-1}$, which corresponds to $F_{\beta}$ = 47.0 T, in excellent agreement with the Fourier analysis.
The extrapolation of the line intercepts the $1/B$ = 0 line at $n$ = 0.02(2).
The obtained intercept is much closer to the 2D value (0) than the 3D values ($\pm1/8$) and supports the two dimensionality of the $\beta$ cylinder.
Note that the intercept also confirms that $\phi_B$ = 0

We next examine the $\alpha$ oscillation.
Figure \ref{invBcos}(b) shows the oscillatory resistivities after box-smoothing. 
The box (window) size has been matched to the period of the $\beta$ oscillation so that the $\beta$ oscillation has been suppressed effectively.
Looking at $F_{\alpha}/B = 2$, we notice that the curves show a minimum for $\theta \leqslant 48.3^{\circ}$ but a maximum for $\theta \geqslant 57.4^{\circ}$ up to 71.8$^{\circ}$
(it is difficult to classify the curve for $\theta = 52.6^{\circ}$).
This indicates that the first spin-zero for the $\alpha$ oscillation occurs between 48.3 and 57.4$^{\circ}$.
Let us assume that it is at $\theta = 50^{\circ}$, where $\mu^*$ = 0.596 and hence $g = 1.68(1+2n)$.
For $n \geqslant 1$, $R_s$ changes sign at least once between $\theta$ = 57.4 and 71.8$^{\circ}$, which is inconsistent with the experimental observation.
Hence, $n = 0$, and $R_s > 0$ at $\theta = 0$ ($\pi g \mu^*/2 = 0.321\pi$).
A more general analysis indicates that a $g$ value between 1.41 and 1.73 can explain the experimental observation and confirms $R_s > 0$ at $\theta = 0$.
Since the curve for $\theta = 0$ shows a minimum near $F_{\alpha}/B = 2$ and maxima near $F_{\alpha}/B = 1.5$ and 2.5, $\phi_B = \pi$ for the $\alpha$ oscillation.
That is, the $\alpha$ cylinder is a Fermi surface of Dirac fermions.
(It is difficult to determine whether $\phi_D$ = 0 or $\pm \pi/4$ for the $\alpha$ oscillation: the number of the observed oscillation periods is too small and the resistivity maxima and minima cannot be determined accurately because the oscillation is weak.)

\section{comparison with band-structure calculations}
\begin{figure}[h]
\includegraphics[width=7.7cm]{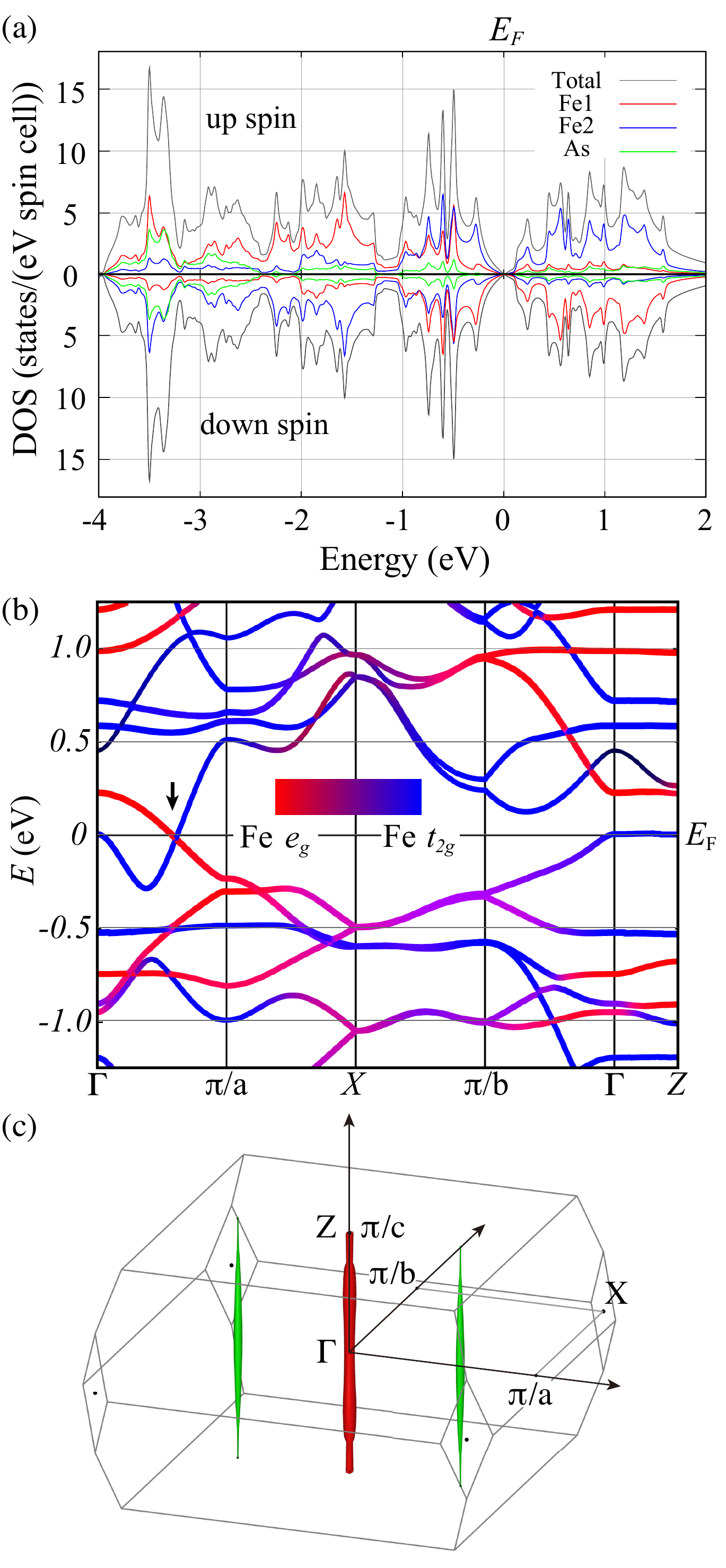}
\caption{\label{band}  Results of the band-structure calculations with spin-orbit coupling.  (a) The total and partial densities of states.  Up-spin is shown above the axis, and down-spin below.  Fe1 and Fe2 refer to iron sites belonging to different magnetic sublattices (see Fig. \ref{struct}).  (b) Band structure near the Fermi level.  The vertical arrow indicates the Dirac point.  The color coding is based on orbital weights as explained by the inset, where $e_g$ refers to $d_{Z^2}+d_{X^2-Y^2}$, while $t_{2g}$ $d_{XZ}+d_{YZ}+d_{XY}$.  (c) Fermi surface in the antiferromagnetic Brillouin zone.}   
\end{figure}

We now consider results of our band-structure calculations.
Figure \ref{band} shows the calculated density of states, band structure, and Fermi surface.
Those results are reminiscent of similar results for the antiferromagnetic state of LaFeAsO \cite{Yin08PRL}.
The density of states shows that the Fermi level is located at the bottom of a deep pseudo gap.
The density of states at $E_F$ is of the order of 0.1 states/eV per spin per cell.
The difference in the occupied states below $E_F$ between up- and down-spin at the Fe1 or Fe2 site gives an antiferromagnetic moment of 2.0 $\mu_B$, which is much larger than the experimental value of 0.49(5) $\mu_B$ \cite{Xiao09PRB_CaFeAsF}.
We realize that the significant difference in the Fe1 (or Fe2) partial density of states between up- and down-spin is not restricted to near $E_F$ but exists even at 4 eV below $E_F$.
This is very different from a textbook picture of a spin-density wave where the spin polarization is expected to occur only near $E_F$ and indicates that the antiferromagnetic order involves states far below $E_F$ as was pointed out early on in e.g. \cite{Johannes09PRB}.

The calculated band structure [Fig.~\ref{band}(b)] shows that two bands cross $E_F$, giving rise to a hole cylinder at the center of the Brillouin zone $\Gamma$ and two symmetry-related electron cylinders off $\Gamma$ toward the antiferromagnetic $a$ direction [Fig.~\ref{band}(c)].
We note that the electron cylinder encloses a Dirac line: 
See the band structure along the line between $\Gamma$ and `$\pi/a$'.
An $e_g$ and a $t_{2g}$ band cross immediately below $E_F$ without spin-orbit coupling.
Spin-orbit coupling induces a gap of about 5 meV, which is too small to be visible in Fig.~\ref{band}(b).
We also note that the two $E_F$-crossing bands are very flat along the line $\Gamma$Z, which indicates a highly two-dimensional electronic state and is consistent with the large anisotropy of the resistivity.
This contrasts with the case of antiferromagnetic state of BaFe$_2$As$_2$, where the band structure is three dimensional, producing closed Fermi pockets \cite{Shimojima10PRL, Terashima11PRL}.

We attribute the $\alpha$ and $\beta$ frequencies to the electron and hole cylinders, respectively.
The Dirac nature of the $\alpha$ cylinder is thus confirmed theoretically as well.  
Based on this assignment and from the experimental frequencies and effective masses, we can estimate the carrier densities to be $n_h$ = 1.7 $\times 10^{-3}$ holes/Fe and $n_e$ = 1.4 $\times 10^{-3}$ electrons/Fe, and the Sommerfeld coefficient to be 1.3 mJ/(mol K$^2$).
The slight imperfection of the carrier compensation can be ascribed to possible small corrugation of the $\alpha$ cylinder, error in the $\alpha$ frequency, and/or off-stoichiometry.
See table I for other parameters such as Fermi velocity and energy.

\begin{table}
\caption{\label{Tab} Carrier type, quantum-oscillation frequency $F$, effective mass $m^*$, carrier density $n$, Fermi momentum $k_F$, Fermi velocity $v_F$, and effective Fermi energy $E_F$ determined from the quantum-oscillation measurements.  $B\parallel c$ for $F$ and $m^*$.  $n$ for the $\alpha$ cylinder refers to the sum of the carriers of the two cylinders occurring in the Brillouin zone.  To derive $E_F$, a linear dispersion ($E_F = \hbar v_F k_F$) was used for the $\alpha$ cylinder, while a quadratic one ($E_F = \hbar^2 k_F^2 / (2m^*)$) for $\beta$.}
\begin{ruledtabular}
\begin{tabular}{ccc}
FS cylinder & $\alpha$ & $\beta$\\
\hline
carrier type & Dirac electron & normal hole\\
$F$ (T) & 19.6 & 46.7 \\
$m^*/m_e$ & 0.383(4) & 0.92(1) \\
$n$ (10$^{-3}$ per Fe) & 1.4 & 1.7 \\
$n$ (10$^{19}$ cm$^{-3}$) & 2.2 & 2.6\\
$k_F$ (\AA$^{-1}$) & 0.024 & 0.038\\
$v_F$ (10$^4$ m/s) & 7.4 & 4.7\\
$E_F$ (meV) & 12 & 5.9\\
\end{tabular}
\end{ruledtabular}
\end{table}

\section{discussion}

First of all, the standard band-structure calculations explain the Fermi surface in the antiferromagnetic state of CaFeAsF reasonably well despite the fact that they significantly overestimate the antiferromagnetic moment.
This is very similar to what we saw in the case of the antiferromagnetic state of BaFe$_2$As$_2$ \cite{Terashima11PRL}.
Since the antiferromagnetic ordering involves deep states (at least) down to 4 eV below $E_F$, band-energy adjustments of order of 100 meV, which are often employed to improve the agreement between the experimental and calculated Fermi surfaces, cannot resolve the disagreement about the moment.

It is interesting to note here that the Fermi surface in the low-temperature nonmagnetic state of FeSe strikingly differs from that expected from band-structure calculations \cite{Terashima14PRB, Tan13NatMat, Maletz14PRB, Nakayama14PRL, Shimojima14PRB, Audouard15EPL, Watson15PRB, Watson15PRL}: notably, the carrier density is more than one order-of-magnitude smaller than calculated \cite{Terashima14PRB}.
In the case of FeSe, only the structural (nematic) transition occurs without antiferromagnetic ordering \cite{McQueen09PRL}.
The stark contrast between the above antiferromagnetic compounds and FeSe seems to suggest that interactions that make the Fermi surface significantly deviate from band-structure calculations are largely suppressed by antiferromagnetic ordering.

The observed antiferromagnetic metallic state of CaFeAsF is unusual in a sense that the carrier density is as low as 10$^{-3}$ per Fe and that one type of the carriers is Dirac fermions.
One may ask why CaFeAsF does not become an insulator after the antiferromagnetic ordering despite the strong antiferromagnetic interaction involving states far below $E_F$.
The unusual metallic state is actually a decisive demonstration of an early theoretical idea by Ran \textit{et al}. \cite{ Ran09PRB} (see also \cite{Morinari10PRL}).
According to Ran \textit{et al}., CaFeAsF is not allowed to become an insulator.
Because of the topological feature of the iron-pnictide band structures, the antiferromagnetic gap has to have two Dirac nodes (in a 2D approximation).
The nodes are close to but offset from $E_F$, giving rise to FS pockets.

The theory by Ran \textit{et al}. is also applicable to the antiferromagnetic state of $R$FeAsO ($R$ = rare earth) and $A$Fe$_2$As$_2$ ($A$ = Ba, Sr, or Ca).
The existence of Dirac fermions in those compounds has been suggested by magnetotransport \cite{Huynh11PRL, Tanabe11PRB, Pallecchi11PRB}, angle-resolved photoemission spectroscopy \cite{Richard10PRL, Shimojima10PRL, Yi11PNAS}, and optical measurements \cite{Nakajima11PNAS, Chen17PRL}.
The Fermi velocity ($\sim 5 \times 10^4$ m/s) of the Dirac cone in BaFe$_2$As$_2$ estimated from angle-resolved photoemission data \cite{Richard10PRL} is close to that of the present $\alpha$ cylinder (Table I).
However, quantum oscillation measurements so far have not confirmed a nontrivial Berry phase in those compounds \cite{Sebastian08JPCM, Harrison09JPCM, Analytis09PRB, Terashima11PRL, Sutherland11PRB, Graf12PRB, Rosa14PRB, Caglieris17PRB}.
Thus this is the first observation of a nontrivial Berry phase in iron-based superconductor parent compounds.

It would be interesting if one could eliminate the hole cylinder by slightly electron-doping CaFeAsF.
One could investigate transport properties of Dirac fermions.
Further, one might be able to induce topological superconductivity of Dirac fermions.
In the case of BaFe$_2$As$_2$, in addition to the Dirac electron pockets (the $\gamma$ pockets in \cite{Terashima11PRL}), there are hole ($\alpha$) and electron ($\delta$) pockets, and it is difficult to simultaneously remove both of the latter pockets by doping.
Therefore CaFeAsF is best suited to testing those possibilities of Dirac-fermion physics.

We have observed nearly $B$-linear magnetoresistance [Fig.~\ref{sig20T}(a)].
Similar $B$-linear magnetoresistance has been reported for other Dirac-fermion systems such as Bi \cite{Kapitza28rspa, Fuseya15JPSJ}, Cd$_3$As$_2$ \cite{Liang15nmat, He14PRL, Feng15PRB}, Na$_3$Bi \cite{Kushwaha15APLMaterials}, TlBiSSe \cite{Novak15PRB}, and ZrSiS \cite{Singha17PNAS}, and further it has been reported for $R$FeAsO \cite{Pallecchi11PRB, Pallecchi13EPJB} and BaFe$_2$As$_2$ \cite{Huynh11PRL} as well.
It is sometimes analyzed with a model of quantum linear magnetoresistance proposed by Abrikosov \cite{Abrikosov98PRB}.
However, this model is applicable only in the quantum limit.
The present linear magnetoresistance starts at about 0.6 T, which is far smaller than a field to reach the quantum limit.
This observation is similar to those reported in \cite{Liang15nmat, He14PRL, Feng15PRB, Kushwaha15APLMaterials, Novak15PRB, Singha17PNAS}.
Since the linear magnetoresistance appears fairly common among Dirac-fermion systems, it might be rooted in characteristic scattering mechanisms of Dirac fermions in magnetic fields as discussed in \cite{Liang15nmat, Feng15PRB}.

The nonmetallic conductivity above $T_s$ is enigmatic.
Band-structure calculations for the paramagnetic state predict a large Fermi surface similar to that in the paramagnetic state of other 1111 or 122 compounds \cite{Ma17SST}.
On the other hand, it has theoretically been suggested that as the As height, the vertical distance of As from the Fe plane, increases, the Fe 3$d$ band width decreases, leading to enhanced electronic fluctuations and stronger electronic correlations \cite{Vildosola08PRB, Kuroki09PRB}.
The As height in CaFeAsF is 1.41 \AA \cite{Ma15SST}, which is  appreciably larger than 1.32 \AA~in LaFeAsO \cite{Cruz08nature}.
Therefore, because of enhanced fluctuations, CaFeAsF above $T_s$ may be in an incoherent state where the Fermi surface is not well-defined.

We also note that the conductivity becomes metallic below the structural (nematic) transition at $T_s$, not the antiferromagnetic one at $T_N$.
Since no band folding occurs at $T_s$ and the orthorhombic distortion is tiny (0.3\%), almost no change is expected for the calculated Fermi surface at $T_s$.
This may indicate that the metallic conductivity below $T_s$ is owing to the suppression of nematic fluctuations, which in turn may indicate the importance of those fluctuations in driving CaFeAsF into an incoherent state.

It is interesting to note that the conductivity near room temperature becomes metallic with Co doping: the resistivity of Ca(Fe$_{0.88}$Co$_{0.12}$)AsF decreases from room temperature down to about 100 K, where it exhibits a broad and shallow minimum \cite{Xiao09PRB_CaFeAsF, MA16JCG}.
This behavior is very similar to that observed in doped $R$FeAsO, e.g., LaFeAs(O$_{1-x}$F$_x$) with $x$ = 0.03 or 0.05 and Sm(Fe$_{0.92}$Co$_{0.08}$)AsO \cite{Dong08EPL, Kohama08PRB, Zhigadlo12PRB}.

Finally, we mention a very recent report of quantum oscillation measurements on SmFeAsO \cite{Caglieris17PRB}.
Only one frequency was observed for $B \parallel c$, and its angular dependence was not determined.
The reported frequency $F$ = 65(5) T is close to the $\beta$ frequency, but the effective mass 0.5(1) $m_e$ is about half of the $\beta$ mass.

\section{summary}
We have compehensively determined the Fermi surface of antiferromagnetic CaFeAsF: it consists of the $\beta$ hole cylinder at the $\Gamma$ point and a pair of the $\alpha$ Dirac electron cylinders symmetrically located at positions off $\Gamma$ toward the antiferromagnetic direction.
We have proved that a nontrivial Berry phase $\pi$ is associated with the $\alpha$ Dirac cylinders.
The carrier density is extremely small, of the order of 10$^{-3}$ per Fe.
This unusual metallic state is a consequence of the previously studied topological nature of the band structure \cite{ Ran09PRB}.
A nearly linear magnetoresistance is observed as in other Dirac-fermion systems.
The anomalous resistivity increase observed above $\sim$30 T for $B \parallel c$ is likely related to the quantum limit of the $\alpha$ orbit.
The band-structure calculations describe the Fermi surface reasonably well.
The same calculations however largely overestimate the antiferromagnetic moment.
Electron interactions that preserve the Fermi surface have to be sought to resolve this problem.
The nonmetallic conduction observed above $T_s$ is intriguing given that the low-temperature antiferromagnetic state is metallic with a well-defined Fermi surface.
It may suggest that CaFeAsF is in an incoherent state above $T_s$. 
Finally, slight electron-doping can make CaFeAsF a system composed solely of Dirac fermions.
Investigations of transport and other properties there would be fruitful, and moreover Dirac-fermion superconductivity might be induced.

\begin{acknowledgments}
We thank Yinguo Xiao for providing us with the structural parameters for the antiferromagnetic orthorhombic phase of CaFeAsF.
We thank Satoru Matsuishi, Hiroyuki Yamase, Tomoaki Agatsuma, and Yuki Fuseya for valuable discussions.

This work was supported by JSPS KAKENHI Grants No. JP17K05556, No. JP17J06088, No. JP16H04021, No. JP16H01081, the Youth Innovation Promotion Association of the Chinese Academy of Sciences (No. 2015187), and the National Natural Science Foundation of China (No. 11574338).
A portion of this work was performed at the National High Magnetic Field Laboratory, which is supported by National Science Foundation Cooperative Agreement No. DMR-1157490 and the State of Florida.
\end{acknowledgments}


\begin{thebibliography}{76}%
\makeatletter
\providecommand \@ifxundefined [1]{%
 \@ifx{#1\undefined}
}%
\providecommand \@ifnum [1]{%
 \ifnum #1\expandafter \@firstoftwo
 \else \expandafter \@secondoftwo
 \fi
}%
\providecommand \@ifx [1]{%
 \ifx #1\expandafter \@firstoftwo
 \else \expandafter \@secondoftwo
 \fi
}%
\providecommand \natexlab [1]{#1}%
\providecommand \enquote  [1]{``#1''}%
\providecommand \bibnamefont  [1]{#1}%
\providecommand \bibfnamefont [1]{#1}%
\providecommand \citenamefont [1]{#1}%
\providecommand \href@noop [0]{\@secondoftwo}%
\providecommand \href [0]{\begingroup \@sanitize@url \@href}%
\providecommand \@href[1]{\@@startlink{#1}\@@href}%
\providecommand \@@href[1]{\endgroup#1\@@endlink}%
\providecommand \@sanitize@url [0]{\catcode `\\12\catcode `\$12\catcode
  `\&12\catcode `\#12\catcode `\^12\catcode `\_12\catcode `\%12\relax}%
\providecommand \@@startlink[1]{}%
\providecommand \@@endlink[0]{}%
\providecommand \url  [0]{\begingroup\@sanitize@url \@url }%
\providecommand \@url [1]{\endgroup\@href {#1}{\urlprefix }}%
\providecommand \urlprefix  [0]{URL }%
\providecommand \Eprint [0]{\href }%
\providecommand \doibase [0]{http://dx.doi.org/}%
\providecommand \selectlanguage [0]{\@gobble}%
\providecommand \bibinfo  [0]{\@secondoftwo}%
\providecommand \bibfield  [0]{\@secondoftwo}%
\providecommand \translation [1]{[#1]}%
\providecommand \BibitemOpen [0]{}%
\providecommand \bibitemStop [0]{}%
\providecommand \bibitemNoStop [0]{.\EOS\space}%
\providecommand \EOS [0]{\spacefactor3000\relax}%
\providecommand \BibitemShut  [1]{\csname bibitem#1\endcsname}%
\let\auto@bib@innerbib\@empty
\bibitem [{\citenamefont {Kamihara}\ \emph {et~al.}(2008)\citenamefont
  {Kamihara}, \citenamefont {Watanabe}, \citenamefont {Hirano},\ and\
  \citenamefont {Hosono}}]{Kamihara08JACS}%
  \BibitemOpen
  \bibfield  {author} {\bibinfo {author} {\bibfnamefont {Y.}~\bibnamefont
  {Kamihara}}, \bibinfo {author} {\bibfnamefont {T.}~\bibnamefont {Watanabe}},
  \bibinfo {author} {\bibfnamefont {M.}~\bibnamefont {Hirano}}, \ and\ \bibinfo
  {author} {\bibfnamefont {H.}~\bibnamefont {Hosono}},\ }\bibfield  {title}
  {\enquote {\bibinfo {title} {{Iron-Based Layered Superconductor
  La[O$_{1-x}$F$_x$]FeAs ($x$ = 0.05--0.12) with $T_c$ = 26 K}},}\ }\href
  {\doibase 10.1021/ja800073m} {\bibfield  {journal} {\bibinfo  {journal} {J.
  Am. Chem. Soc.}\ }\textbf {\bibinfo {volume} {130}},\ \bibinfo {pages} {3296}
  (\bibinfo {year} {2008})}\BibitemShut {NoStop}%
\bibitem [{\citenamefont {Kito}\ \emph {et~al.}(2008)\citenamefont {Kito},
  \citenamefont {Eisaki},\ and\ \citenamefont {Iyo}}]{Kito08JPSJ}%
  \BibitemOpen
  \bibfield  {author} {\bibinfo {author} {\bibfnamefont {H.}~\bibnamefont
  {Kito}}, \bibinfo {author} {\bibfnamefont {H.}~\bibnamefont {Eisaki}}, \ and\
  \bibinfo {author} {\bibfnamefont {A.}~\bibnamefont {Iyo}},\ }\bibfield
  {title} {\enquote {\bibinfo {title} {{Superconductivity at 54\,K in F-Free
  NdFeAsO$_{1-y}$}},}\ }\href {\doibase 10.1143/JPSJ.77.063707} {\bibfield
  {journal} {\bibinfo  {journal} {J. Phys. Soc. Jpn.}\ }\textbf {\bibinfo
  {volume} {77}},\ \bibinfo {pages} {063707} (\bibinfo {year}
  {2008})}\BibitemShut {NoStop}%
\bibitem [{\citenamefont {Ren}\ \emph {et~al.}(2008)\citenamefont {Ren},
  \citenamefont {Lu}, \citenamefont {Yang}, \citenamefont {Yi}, \citenamefont
  {Shen}, \citenamefont {Zheng-Cai}, \citenamefont {Guang-Can}, \citenamefont
  {Dong}, \citenamefont {Sun}, \citenamefont {Zhou},\ and\ \citenamefont
  {Zhao}}]{Ren08CPL}%
  \BibitemOpen
  \bibfield  {author} {\bibinfo {author} {\bibfnamefont {Z.-A.}\ \bibnamefont
  {Ren}}, \bibinfo {author} {\bibfnamefont {W.}~\bibnamefont {Lu}}, \bibinfo
  {author} {\bibfnamefont {J.}~\bibnamefont {Yang}}, \bibinfo {author}
  {\bibfnamefont {W.}~\bibnamefont {Yi}}, \bibinfo {author} {\bibfnamefont
  {X.-L.}\ \bibnamefont {Shen}}, \bibinfo {author} {\bibnamefont {Zheng-Cai}},
  \bibinfo {author} {\bibfnamefont {C.}~\bibnamefont {Guang-Can}}, \bibinfo
  {author} {\bibfnamefont {X.-L.}\ \bibnamefont {Dong}}, \bibinfo {author}
  {\bibfnamefont {L.-L.}\ \bibnamefont {Sun}}, \bibinfo {author} {\bibfnamefont
  {F.}~\bibnamefont {Zhou}}, \ and\ \bibinfo {author} {\bibfnamefont {Z.-X.}\
  \bibnamefont {Zhao}},\ }\bibfield  {title} {\enquote {\bibinfo {title}
  {{Superconductivity at 55 K in Iron-Based F-Doped Layered Quaternary Compound
  Sm[O$_{1- x}$F$_x$] FeAs}},}\ }\href
  {http://stacks.iop.org/0256-307X/25/i=6/a=080} {\bibfield  {journal}
  {\bibinfo  {journal} {Chin. Phys. Lett.}\ }\textbf {\bibinfo {volume} {25}},\
  \bibinfo {pages} {2215} (\bibinfo {year} {2008})}\BibitemShut {NoStop}%
\bibitem [{\citenamefont {Yang}\ \emph {et~al.}(2008)\citenamefont {Yang},
  \citenamefont {Li}, \citenamefont {Lu}, \citenamefont {Yi}, \citenamefont
  {Shen}, \citenamefont {Ren}, \citenamefont {Che}, \citenamefont {Dong},
  \citenamefont {Sun}, \citenamefont {Zhou},\ and\ \citenamefont
  {Zhao}}]{Yang08SST}%
  \BibitemOpen
  \bibfield  {author} {\bibinfo {author} {\bibfnamefont {J.}~\bibnamefont
  {Yang}}, \bibinfo {author} {\bibfnamefont {Z.-C.}\ \bibnamefont {Li}},
  \bibinfo {author} {\bibfnamefont {W.}~\bibnamefont {Lu}}, \bibinfo {author}
  {\bibfnamefont {W.}~\bibnamefont {Yi}}, \bibinfo {author} {\bibfnamefont
  {X.-L.}\ \bibnamefont {Shen}}, \bibinfo {author} {\bibfnamefont {Z.-A.}\
  \bibnamefont {Ren}}, \bibinfo {author} {\bibfnamefont {G.-C.}\ \bibnamefont
  {Che}}, \bibinfo {author} {\bibfnamefont {X.-L.}\ \bibnamefont {Dong}},
  \bibinfo {author} {\bibfnamefont {L.-L.}\ \bibnamefont {Sun}}, \bibinfo
  {author} {\bibfnamefont {F.}~\bibnamefont {Zhou}}, \ and\ \bibinfo {author}
  {\bibfnamefont {Z.-X.}\ \bibnamefont {Zhao}},\ }\bibfield  {title} {\enquote
  {\bibinfo {title} {{Superconductivity at 53.5~K in GdFeAsO$_{1-\delta}$}},}\
  }\href {http://stacks.iop.org/0953-2048/21/082001} {\bibfield  {journal}
  {\bibinfo  {journal} {Supercond. Sci. Technol.}\ }\textbf {\bibinfo {volume}
  {21}},\ \bibinfo {pages} {082001} (\bibinfo {year} {2008})}\BibitemShut
  {NoStop}%
\bibitem [{\citenamefont {Wang}\ \emph
  {et~al.}(2008{\natexlab{a}})\citenamefont {Wang}, \citenamefont {Li},
  \citenamefont {Chi}, \citenamefont {Zhu}, \citenamefont {Ren}, \citenamefont
  {Li}, \citenamefont {Wang}, \citenamefont {Lin}, \citenamefont {Luo},
  \citenamefont {Jiang}, \citenamefont {Xu}, \citenamefont {Cao},\ and\
  \citenamefont {Xu}}]{Wang08EPL}%
  \BibitemOpen
  \bibfield  {author} {\bibinfo {author} {\bibfnamefont {C.}~\bibnamefont
  {Wang}}, \bibinfo {author} {\bibfnamefont {L.}~\bibnamefont {Li}}, \bibinfo
  {author} {\bibfnamefont {S.}~\bibnamefont {Chi}}, \bibinfo {author}
  {\bibfnamefont {Z.}~\bibnamefont {Zhu}}, \bibinfo {author} {\bibfnamefont
  {Z.}~\bibnamefont {Ren}}, \bibinfo {author} {\bibfnamefont {Y.}~\bibnamefont
  {Li}}, \bibinfo {author} {\bibfnamefont {Y.}~\bibnamefont {Wang}}, \bibinfo
  {author} {\bibfnamefont {X.}~\bibnamefont {Lin}}, \bibinfo {author}
  {\bibfnamefont {Y.}~\bibnamefont {Luo}}, \bibinfo {author} {\bibfnamefont
  {S.}~\bibnamefont {Jiang}}, \bibinfo {author} {\bibfnamefont
  {X.}~\bibnamefont {Xu}}, \bibinfo {author} {\bibfnamefont {G.}~\bibnamefont
  {Cao}}, \ and\ \bibinfo {author} {\bibfnamefont {Z.}~\bibnamefont {Xu}},\
  }\bibfield  {title} {\enquote {\bibinfo {title} {{Thorium-Doping-Induced
  Superconductivity up to 56 K in Gd$_{1-x}$Th$_x$FeAsO}},}\ }\href
  {http://stacks.iop.org/0295-5075/83/67006} {\bibfield  {journal} {\bibinfo
  {journal} {EPL}\ }\textbf {\bibinfo {volume} {83}},\ \bibinfo {pages} {67006}
  (\bibinfo {year} {2008}{\natexlab{a}})}\BibitemShut {NoStop}%
\bibitem [{\citenamefont {Rotter}\ \emph {et~al.}(2008)\citenamefont {Rotter},
  \citenamefont {Tegel},\ and\ \citenamefont {Johrendt}}]{Rotter08PRL}%
  \BibitemOpen
  \bibfield  {author} {\bibinfo {author} {\bibfnamefont {M.}~\bibnamefont
  {Rotter}}, \bibinfo {author} {\bibfnamefont {M.}~\bibnamefont {Tegel}}, \
  and\ \bibinfo {author} {\bibfnamefont {D.}~\bibnamefont {Johrendt}},\
  }\bibfield  {title} {\enquote {\bibinfo {title} {{Superconductivity at 38 K
  in the Iron Arsenide (Ba$_{1-x}$K$_x$)Fe$_2$As$_2$}},}\ }\href {\doibase
  10.1103/PhysRevLett.101.107006} {\bibfield  {journal} {\bibinfo  {journal}
  {Phys. Rev. Lett.}\ }\textbf {\bibinfo {volume} {101}},\ \bibinfo {eid}
  {107006} (\bibinfo {year} {2008})}\BibitemShut {NoStop}%
\bibitem [{\citenamefont {Sasmal}\ \emph {et~al.}(2008)\citenamefont {Sasmal},
  \citenamefont {Lv}, \citenamefont {Lorenz}, \citenamefont {Guloy},
  \citenamefont {Chen}, \citenamefont {Xue},\ and\ \citenamefont
  {Chu}}]{Sasmal08PRL}%
  \BibitemOpen
  \bibfield  {author} {\bibinfo {author} {\bibfnamefont {K.}~\bibnamefont
  {Sasmal}}, \bibinfo {author} {\bibfnamefont {B.}~\bibnamefont {Lv}}, \bibinfo
  {author} {\bibfnamefont {B.}~\bibnamefont {Lorenz}}, \bibinfo {author}
  {\bibfnamefont {A.~M.}\ \bibnamefont {Guloy}}, \bibinfo {author}
  {\bibfnamefont {F.}~\bibnamefont {Chen}}, \bibinfo {author} {\bibfnamefont
  {Y.-Y.}\ \bibnamefont {Xue}}, \ and\ \bibinfo {author} {\bibfnamefont
  {C.-W.}\ \bibnamefont {Chu}},\ }\bibfield  {title} {\enquote {\bibinfo
  {title} {{Superconducting Fe-Based Compounds (A$_{1-x}$Sr$_x$)Fe$_2$As$_2$
  with A = K and Cs with Transition Temperatures up to 37 K}},}\ }\href
  {\doibase 10.1103/PhysRevLett.101.107007} {\bibfield  {journal} {\bibinfo
  {journal} {Phys. Rev. Lett.}\ }\textbf {\bibinfo {volume} {101}},\ \bibinfo
  {eid} {107007} (\bibinfo {year} {2008})}\BibitemShut {NoStop}%
\bibitem [{\citenamefont {Tapp}\ \emph {et~al.}(2008)\citenamefont {Tapp},
  \citenamefont {Tang}, \citenamefont {Lv}, \citenamefont {Sasmal},
  \citenamefont {Lorenz}, \citenamefont {Chu},\ and\ \citenamefont
  {Guloy}}]{Tapp08PRB}%
  \BibitemOpen
  \bibfield  {author} {\bibinfo {author} {\bibfnamefont {J.~H.}\ \bibnamefont
  {Tapp}}, \bibinfo {author} {\bibfnamefont {Z.}~\bibnamefont {Tang}}, \bibinfo
  {author} {\bibfnamefont {B.}~\bibnamefont {Lv}}, \bibinfo {author}
  {\bibfnamefont {K.}~\bibnamefont {Sasmal}}, \bibinfo {author} {\bibfnamefont
  {B.}~\bibnamefont {Lorenz}}, \bibinfo {author} {\bibfnamefont {P.~C.~W.}\
  \bibnamefont {Chu}}, \ and\ \bibinfo {author} {\bibfnamefont {A.~M.}\
  \bibnamefont {Guloy}},\ }\bibfield  {title} {\enquote {\bibinfo {title}
  {{LiFeAs: An Intrinsic FeAs-Based Superconductor with ${T}_{c}=18\text{
  }\text{K}$}},}\ }\href {\doibase 10.1103/PhysRevB.78.060505} {\bibfield
  {journal} {\bibinfo  {journal} {Phys. Rev. B}\ }\textbf {\bibinfo {volume}
  {78}},\ \bibinfo {pages} {060505} (\bibinfo {year} {2008})}\BibitemShut
  {NoStop}%
\bibitem [{\citenamefont {Pitcher}\ \emph {et~al.}(2008)\citenamefont
  {Pitcher}, \citenamefont {Parker}, \citenamefont {Adamson}, \citenamefont
  {Herkelrath}, \citenamefont {Boothroyd}, \citenamefont {Ibberson},
  \citenamefont {Brunelli},\ and\ \citenamefont {Clarke}}]{Pitcher08ChemCom}%
  \BibitemOpen
  \bibfield  {author} {\bibinfo {author} {\bibfnamefont {M.~J.}\ \bibnamefont
  {Pitcher}}, \bibinfo {author} {\bibfnamefont {D.~R.}\ \bibnamefont {Parker}},
  \bibinfo {author} {\bibfnamefont {P.}~\bibnamefont {Adamson}}, \bibinfo
  {author} {\bibfnamefont {S.~J.~C.}\ \bibnamefont {Herkelrath}}, \bibinfo
  {author} {\bibfnamefont {A.~T.}\ \bibnamefont {Boothroyd}}, \bibinfo {author}
  {\bibfnamefont {R.~M.}\ \bibnamefont {Ibberson}}, \bibinfo {author}
  {\bibfnamefont {M.}~\bibnamefont {Brunelli}}, \ and\ \bibinfo {author}
  {\bibfnamefont {S.~J.}\ \bibnamefont {Clarke}},\ }\bibfield  {title}
  {\enquote {\bibinfo {title} {{Structure and Superconductivity of LiFeAs}},}\
  }\href {\doibase 10.1039/B813153H} {\bibfield  {journal} {\bibinfo  {journal}
  {Chem. Commun. (Cambridge)}\ \textbf {\bibinfo {volume}
  {45}},\ \bibinfo {pages} {5918}} (\bibinfo {year}
  {2008})}\BibitemShut {NoStop}%
\bibitem [{\citenamefont {Wang}\ \emph
  {et~al.}(2008{\natexlab{b}})\citenamefont {Wang}, \citenamefont {Liu},
  \citenamefont {Lv}, \citenamefont {Gao}, \citenamefont {Yang}, \citenamefont
  {Yu}, \citenamefont {Li},\ and\ \citenamefont {Jin}}]{Wang08SSC}%
  \BibitemOpen
  \bibfield  {author} {\bibinfo {author} {\bibfnamefont {X.}~\bibnamefont
  {Wang}}, \bibinfo {author} {\bibfnamefont {Q.}~\bibnamefont {Liu}}, \bibinfo
  {author} {\bibfnamefont {Y.}~\bibnamefont {Lv}}, \bibinfo {author}
  {\bibfnamefont {W.}~\bibnamefont {Gao}}, \bibinfo {author} {\bibfnamefont
  {L.}~\bibnamefont {Yang}}, \bibinfo {author} {\bibfnamefont {R.}~\bibnamefont
  {Yu}}, \bibinfo {author} {\bibfnamefont {F.}~\bibnamefont {Li}}, \ and\
  \bibinfo {author} {\bibfnamefont {C.}~\bibnamefont {Jin}},\ }\bibfield
  {title} {\enquote {\bibinfo {title} {{The Superconductivity at 18 K in LiFeAs
  System}},}\ }\href {\doibase DOI: 10.1016/j.ssc.2008.09.057} {\bibfield
  {journal} {\bibinfo  {journal} {Solid State Commun.}\ }\textbf {\bibinfo
  {volume} {148}},\ \bibinfo {pages} {538} (\bibinfo {year}
  {2008}{\natexlab{b}})}\BibitemShut {NoStop}%
\bibitem [{\citenamefont {Hsu}\ \emph {et~al.}(2008)\citenamefont {Hsu},
  \citenamefont {Luo}, \citenamefont {Yeh}, \citenamefont {Chen}, \citenamefont
  {Huang}, \citenamefont {Wu}, \citenamefont {Lee}, \citenamefont {Huang},
  \citenamefont {Chu}, \citenamefont {Yan},\ and\ \citenamefont
  {Wu}}]{Hsu08PNAS}%
  \BibitemOpen
  \bibfield  {author} {\bibinfo {author} {\bibfnamefont {F.-C.}\ \bibnamefont
  {Hsu}}, \bibinfo {author} {\bibfnamefont {J.-Y.}\ \bibnamefont {Luo}},
  \bibinfo {author} {\bibfnamefont {K.-W.}\ \bibnamefont {Yeh}}, \bibinfo
  {author} {\bibfnamefont {T.-K.}\ \bibnamefont {Chen}}, \bibinfo {author}
  {\bibfnamefont {T.-W.}\ \bibnamefont {Huang}}, \bibinfo {author}
  {\bibfnamefont {P.~M.}\ \bibnamefont {Wu}}, \bibinfo {author} {\bibfnamefont
  {Y.-C.}\ \bibnamefont {Lee}}, \bibinfo {author} {\bibfnamefont {Y.-L.}\
  \bibnamefont {Huang}}, \bibinfo {author} {\bibfnamefont {Y.-Y.}\ \bibnamefont
  {Chu}}, \bibinfo {author} {\bibfnamefont {D.-C.}\ \bibnamefont {Yan}}, \ and\
  \bibinfo {author} {\bibfnamefont {M.-K.}\ \bibnamefont {Wu}},\ }\bibfield
  {title} {\enquote {\bibinfo {title} {{Superconductivity in the PbO-Type
  Structure {$\alpha$}-FeSe}},}\ }\href {\doibase 10.1073/pnas.0807325105}
  {\bibfield  {journal} {\bibinfo  {journal} {Proc. Nat. Acad. Sci. U. S. A.}\
  }\textbf {\bibinfo {volume} {105}},\ \bibinfo {pages} {14262} (\bibinfo
  {year} {2008})}\BibitemShut {NoStop}%
\bibitem [{\citenamefont {Matsuishi}\ \emph {et~al.}(2008)\citenamefont
  {Matsuishi}, \citenamefont {Inoue}, \citenamefont {Nomura}, \citenamefont
  {Yanagi}, \citenamefont {Hirano},\ and\ \citenamefont
  {Hosono}}]{Matsuishi08JACS}%
  \BibitemOpen
  \bibfield  {author} {\bibinfo {author} {\bibfnamefont {S.}~\bibnamefont
  {Matsuishi}}, \bibinfo {author} {\bibfnamefont {Y.}~\bibnamefont {Inoue}},
  \bibinfo {author} {\bibfnamefont {T.}~\bibnamefont {Nomura}}, \bibinfo
  {author} {\bibfnamefont {H.}~\bibnamefont {Yanagi}}, \bibinfo {author}
  {\bibfnamefont {M.}~\bibnamefont {Hirano}}, \ and\ \bibinfo {author}
  {\bibfnamefont {H.}~\bibnamefont {Hosono}},\ }\bibfield  {title} {\enquote
  {\bibinfo {title} {{Superconductivity Induced by Co-Doping in Quaternary
  Fluoroarsenide CaFeAsF}},}\ }\href {\doibase 10.1021/ja806357j} {\bibfield
  {journal} {\bibinfo  {journal} {J. Am. Chem. Soc.}\ }\textbf {\bibinfo
  {volume} {130}},\ \bibinfo {pages} {14428} (\bibinfo {year}
  {2008})}\BibitemShut {NoStop}%
\bibitem [{\citenamefont {Tegel}\ \emph {et~al.}(2008)\citenamefont {Tegel},
  \citenamefont {Johansson}, \citenamefont {Wei{\ss}}, \citenamefont
  {Schellenberg}, \citenamefont {Hermes}, \citenamefont {P{\"o}ttgen},\ and\
  \citenamefont {Johrendt}}]{Tegel08EPL}%
  \BibitemOpen
  \bibfield  {author} {\bibinfo {author} {\bibfnamefont {M.}~\bibnamefont
  {Tegel}}, \bibinfo {author} {\bibfnamefont {S.}~\bibnamefont {Johansson}},
  \bibinfo {author} {\bibfnamefont {V.}~\bibnamefont {Wei{\ss}}}, \bibinfo
  {author} {\bibfnamefont {I.}~\bibnamefont {Schellenberg}}, \bibinfo {author}
  {\bibfnamefont {W.}~\bibnamefont {Hermes}}, \bibinfo {author} {\bibfnamefont
  {R.}~\bibnamefont {P{\"o}ttgen}}, \ and\ \bibinfo {author} {\bibfnamefont
  {D.}~\bibnamefont {Johrendt}},\ }\bibfield  {title} {\enquote {\bibinfo
  {title} {{Synthesis, Crystal Structure and Spin-Density-Wave Anomaly of the
  Iron Arsenide-Fluoride SrFeAsF}},}\ }\href
  {http://stacks.iop.org/0295-5075/84/i=6/a=67007} {\bibfield  {journal}
  {\bibinfo  {journal} {EPL}\ }\textbf {\bibinfo {volume} {84}},\ \bibinfo
  {pages} {67007} (\bibinfo {year} {2008})}\BibitemShut {NoStop}%
\bibitem [{\citenamefont {Han}\ \emph {et~al.}(2008)\citenamefont {Han},
  \citenamefont {Zhu}, \citenamefont {Mu}, \citenamefont {Cheng},\ and\
  \citenamefont {Wen}}]{Han08PRB}%
  \BibitemOpen
  \bibfield  {author} {\bibinfo {author} {\bibfnamefont {F.}~\bibnamefont
  {Han}}, \bibinfo {author} {\bibfnamefont {X.}~\bibnamefont {Zhu}}, \bibinfo
  {author} {\bibfnamefont {G.}~\bibnamefont {Mu}}, \bibinfo {author}
  {\bibfnamefont {P.}~\bibnamefont {Cheng}}, \ and\ \bibinfo {author}
  {\bibfnamefont {H.-H.}\ \bibnamefont {Wen}},\ }\bibfield  {title} {\enquote
  {\bibinfo {title} {{SrFeAsF as a Parent Compound for Iron Pnictide
  Superconductors}},}\ }\href {\doibase 10.1103/PhysRevB.78.180503} {\bibfield
  {journal} {\bibinfo  {journal} {Phys. Rev. B}\ }\textbf {\bibinfo {volume}
  {78}},\ \bibinfo {pages} {180503} (\bibinfo {year} {2008})}\BibitemShut
  {NoStop}%
\bibitem [{\citenamefont {Shein}\ and\ \citenamefont
  {Ivanovskii}(2008)}]{Shein08JETPLett}%
  \BibitemOpen
  \bibfield  {author} {\bibinfo {author} {\bibfnamefont {I.~R.}\ \bibnamefont
  {Shein}}\ and\ \bibinfo {author} {\bibfnamefont {A.~L.}\ \bibnamefont
  {Ivanovskii}},\ }\bibfield  {title} {\enquote {\bibinfo {title} {{Band
  Structure of SrFeAsF and CaFeAsF---the Base Phases of a New Group of
  Oxygen-Free FeAs Superconductors}},}\ }\href {\doibase
  10.1134/S0021364008220104} {\bibfield  {journal} {\bibinfo  {journal} {JETP
  Lett.}\ }\textbf {\bibinfo {volume} {88}},\ \bibinfo {pages} {683} (\bibinfo
  {year} {2008})}\BibitemShut {NoStop}%
\bibitem [{\citenamefont {Xiao}\ \emph {et~al.}(2009)\citenamefont {Xiao},
  \citenamefont {Su}, \citenamefont {Mittal}, \citenamefont {Chatterji},
  \citenamefont {Hansen}, \citenamefont {Kumar}, \citenamefont {Matsuishi},
  \citenamefont {Hosono},\ and\ \citenamefont {Brueckel}}]{Xiao09PRB_CaFeAsF}%
  \BibitemOpen
  \bibfield  {author} {\bibinfo {author} {\bibfnamefont {Y.}~\bibnamefont
  {Xiao}}, \bibinfo {author} {\bibfnamefont {Y.}~\bibnamefont {Su}}, \bibinfo
  {author} {\bibfnamefont {R.}~\bibnamefont {Mittal}}, \bibinfo {author}
  {\bibfnamefont {T.}~\bibnamefont {Chatterji}}, \bibinfo {author}
  {\bibfnamefont {T.}~\bibnamefont {Hansen}}, \bibinfo {author} {\bibfnamefont
  {C.~M.~N.}\ \bibnamefont {Kumar}}, \bibinfo {author} {\bibfnamefont
  {S.}~\bibnamefont {Matsuishi}}, \bibinfo {author} {\bibfnamefont
  {H.}~\bibnamefont {Hosono}}, \ and\ \bibinfo {author} {\bibfnamefont
  {T.}~\bibnamefont {Brueckel}},\ }\bibfield  {title} {\enquote {\bibinfo
  {title} {{Magnetic Order in the
  ${\text{CaFe}}_{1\ensuremath{-}x}{\text{Co}}_{x}\text{AsF}$
  $(x=0.00,0.06,0.12)$ Superconducting Compounds}},}\ }\href {\doibase
  10.1103/PhysRevB.79.060504} {\bibfield  {journal} {\bibinfo  {journal} {Phys.
  Rev. B}\ }\textbf {\bibinfo {volume} {79}},\ \bibinfo {pages} {060504}
  (\bibinfo {year} {2009})}\BibitemShut {NoStop}%
\bibitem [{\citenamefont {Wu}\ \emph {et~al.}(2009)\citenamefont {Wu},
  \citenamefont {Xie}, \citenamefont {Chen}, \citenamefont {Zhong},
  \citenamefont {Liu}, \citenamefont {Shi}, \citenamefont {Li}, \citenamefont
  {Wang}, \citenamefont {Wu}, \citenamefont {Yan}, \citenamefont {Ying},\ and\
  \citenamefont {Chen}}]{Wu09JPCM}%
  \BibitemOpen
  \bibfield  {author} {\bibinfo {author} {\bibfnamefont {G.}~\bibnamefont
  {Wu}}, \bibinfo {author} {\bibfnamefont {Y.~L.}\ \bibnamefont {Xie}},
  \bibinfo {author} {\bibfnamefont {H.}~\bibnamefont {Chen}}, \bibinfo {author}
  {\bibfnamefont {M.}~\bibnamefont {Zhong}}, \bibinfo {author} {\bibfnamefont
  {R.~H.}\ \bibnamefont {Liu}}, \bibinfo {author} {\bibfnamefont {B.~C.}\
  \bibnamefont {Shi}}, \bibinfo {author} {\bibfnamefont {Q.~J.}\ \bibnamefont
  {Li}}, \bibinfo {author} {\bibfnamefont {X.~F.}\ \bibnamefont {Wang}},
  \bibinfo {author} {\bibfnamefont {T.}~\bibnamefont {Wu}}, \bibinfo {author}
  {\bibfnamefont {Y.~J.}\ \bibnamefont {Yan}}, \bibinfo {author} {\bibfnamefont
  {J.~J.}\ \bibnamefont {Ying}}, \ and\ \bibinfo {author} {\bibfnamefont
  {X.~H.}\ \bibnamefont {Chen}},\ }\bibfield  {title} {\enquote {\bibinfo
  {title} {{Superconductivity at 56 K in Samarium-Doped SrFeAsF}},}\ }\href
  {http://stacks.iop.org/0953-8984/21/i=14/a=142203} {\bibfield  {journal}
  {\bibinfo  {journal} {J. Phys.: Condens. Matter}\ }\textbf {\bibinfo {volume}
  {21}},\ \bibinfo {pages} {142203} (\bibinfo {year} {2009})}\BibitemShut
  {NoStop}%
\bibitem [{\citenamefont {Cheng}\ \emph {et~al.}(2009)\citenamefont {Cheng},
  \citenamefont {Shen}, \citenamefont {Mu}, \citenamefont {Zhu}, \citenamefont
  {Han}, \citenamefont {Zeng},\ and\ \citenamefont {Wen}}]{Cheng09EPL}%
  \BibitemOpen
  \bibfield  {author} {\bibinfo {author} {\bibfnamefont {P.}~\bibnamefont
  {Cheng}}, \bibinfo {author} {\bibfnamefont {B.}~\bibnamefont {Shen}},
  \bibinfo {author} {\bibfnamefont {G.}~\bibnamefont {Mu}}, \bibinfo {author}
  {\bibfnamefont {X.}~\bibnamefont {Zhu}}, \bibinfo {author} {\bibfnamefont
  {F.}~\bibnamefont {Han}}, \bibinfo {author} {\bibfnamefont {B.}~\bibnamefont
  {Zeng}}, \ and\ \bibinfo {author} {\bibfnamefont {H.-H.}\ \bibnamefont
  {Wen}},\ }\bibfield  {title} {\enquote {\bibinfo {title} {{High-$T_c$
  Superconductivity Induced by Doping Rare-Earth Elements into CaFeAsF}},}\
  }\href {\doibase 10.1209/0295-5075/85/67003} {\bibfield  {journal} {\bibinfo
  {journal} {EPL}\ }\textbf {\bibinfo {volume} {85}},\ \bibinfo {pages} {67003}
  (\bibinfo {year} {2009})}\BibitemShut {NoStop}%
\bibitem [{\citenamefont {Shlyk}\ \emph {et~al.}(2014)\citenamefont {Shlyk},
  \citenamefont {Wolff}, \citenamefont {Bischoff}, \citenamefont {Rose},
  \citenamefont {Schleid},\ and\ \citenamefont {Niewa}}]{Shlyk14SST}%
  \BibitemOpen
  \bibfield  {author} {\bibinfo {author} {\bibfnamefont {L.}~\bibnamefont
  {Shlyk}}, \bibinfo {author} {\bibfnamefont {K.~K.}\ \bibnamefont {Wolff}},
  \bibinfo {author} {\bibfnamefont {M.}~\bibnamefont {Bischoff}}, \bibinfo
  {author} {\bibfnamefont {E.}~\bibnamefont {Rose}}, \bibinfo {author}
  {\bibfnamefont {T.}~\bibnamefont {Schleid}}, \ and\ \bibinfo {author}
  {\bibfnamefont {R.}~\bibnamefont {Niewa}},\ }\bibfield  {title} {\enquote
  {\bibinfo {title} {{Crystal Structure and Superconducting Properties of
  Hole-Doped Ca$_{0.89}$Na$_{0.11}$FFeAs Single Crystals}},}\ }\href
  {http://stacks.iop.org/0953-2048/27/i=4/a=044011} {\bibfield  {journal}
  {\bibinfo  {journal} {Supercond. Sci. Technol.}\ }\textbf {\bibinfo {volume}
  {27}},\ \bibinfo {pages} {044011} (\bibinfo {year} {2014})}\BibitemShut
  {NoStop}%
\bibitem [{\citenamefont {Tao}\ \emph {et~al.}(2014)\citenamefont {Tao},
  \citenamefont {Li}, \citenamefont {Zhu}, \citenamefont {Yang},\ and\
  \citenamefont {Wen}}]{Tao14SciChina}%
  \BibitemOpen
  \bibfield  {author} {\bibinfo {author} {\bibfnamefont {J.}~\bibnamefont
  {Tao}}, \bibinfo {author} {\bibfnamefont {S.}~\bibnamefont {Li}}, \bibinfo
  {author} {\bibfnamefont {X.}~\bibnamefont {Zhu}}, \bibinfo {author}
  {\bibfnamefont {H.}~\bibnamefont {Yang}}, \ and\ \bibinfo {author}
  {\bibfnamefont {H.-H.}\ \bibnamefont {Wen}},\ }\bibfield  {title} {\enquote
  {\bibinfo {title} {{Growth and Transport Properties of CaFeAsF$_{1-x}$ Single
  Crystals}},}\ }\href {\doibase 10.1007/s11433-014-5422-4} {\bibfield
  {journal} {\bibinfo  {journal} {Sci. Chin.: Phys. Mecha. Astro.}\ }\textbf
  {\bibinfo {volume} {57}},\ \bibinfo {pages} {632} (\bibinfo {year}
  {2014})}\BibitemShut {NoStop}%
\bibitem [{\citenamefont {Ma}\ \emph {et~al.}(2015)\citenamefont {Ma},
  \citenamefont {Zhang}, \citenamefont {Gao}, \citenamefont {Hu}, \citenamefont
  {Ji}, \citenamefont {Mu}, \citenamefont {Huang},\ and\ \citenamefont
  {Xie}}]{Ma15SST}%
  \BibitemOpen
  \bibfield  {author} {\bibinfo {author} {\bibfnamefont {Y.}~\bibnamefont
  {Ma}}, \bibinfo {author} {\bibfnamefont {H.}~\bibnamefont {Zhang}}, \bibinfo
  {author} {\bibfnamefont {B.}~\bibnamefont {Gao}}, \bibinfo {author}
  {\bibfnamefont {K.}~\bibnamefont {Hu}}, \bibinfo {author} {\bibfnamefont
  {Q.}~\bibnamefont {Ji}}, \bibinfo {author} {\bibfnamefont {G.}~\bibnamefont
  {Mu}}, \bibinfo {author} {\bibfnamefont {F.}~\bibnamefont {Huang}}, \ and\
  \bibinfo {author} {\bibfnamefont {X.}~\bibnamefont {Xie}},\ }\bibfield
  {title} {\enquote {\bibinfo {title} {{Growth and Characterization of
  Millimeter-Sized Single Crystals of CaFeAsF}},}\ }\href
  {http://stacks.iop.org/0953-2048/28/i=8/a=085008} {\bibfield  {journal}
  {\bibinfo  {journal} {Supercond. Sci. Technol.}\ }\textbf {\bibinfo {volume}
  {28}},\ \bibinfo {pages} {085008} (\bibinfo {year} {2015})}\BibitemShut
  {NoStop}%
\bibitem [{\citenamefont {Blaha}\ \emph {et~al.}(2001)\citenamefont {Blaha},
  \citenamefont {Schwarz}, \citenamefont {Madsen}, \citenamefont {Kvasnicka},\
  and\ \citenamefont {Luitz}}]{WIEN2K}%
  \BibitemOpen
  \bibfield  {author} {\bibinfo {author} {\bibfnamefont {P.}~\bibnamefont
  {Blaha}}, \bibinfo {author} {\bibfnamefont {K.}~\bibnamefont {Schwarz}},
  \bibinfo {author} {\bibfnamefont {G.~K.~H.}\ \bibnamefont {Madsen}}, \bibinfo
  {author} {\bibfnamefont {D.}~\bibnamefont {Kvasnicka}}, \ and\ \bibinfo
  {author} {\bibfnamefont {J.}~\bibnamefont {Luitz}},\ }\href@noop {} {\emph
  {\bibinfo {title} {{WIEN2k}}}},\ edited by\ \bibinfo {editor} {\bibfnamefont
  {K.}~\bibnamefont {Schwarz}}\ (\bibinfo  {publisher} {Technische
  Universit\"at Wien},\ \bibinfo {address} {Austria},\ \bibinfo {year}
  {2001})\BibitemShut {NoStop}%
\bibitem [{\citenamefont {Perdew}\ \emph {et~al.}(1996)\citenamefont {Perdew},
  \citenamefont {Burke},\ and\ \citenamefont {Ernzerhof}}]{Perdew96PRL}%
  \BibitemOpen
  \bibfield  {author} {\bibinfo {author} {\bibfnamefont {J.~P.}\ \bibnamefont
  {Perdew}}, \bibinfo {author} {\bibfnamefont {K.}~\bibnamefont {Burke}}, \
  and\ \bibinfo {author} {\bibfnamefont {M.}~\bibnamefont {Ernzerhof}},\
  }\bibfield  {title} {\enquote {\bibinfo {title} {{Generalized Gradient
  Approximation Made Simple}},}\ }\href {\doibase 10.1103/PhysRevLett.77.3865}
  {\bibfield  {journal} {\bibinfo  {journal} {Phys. Rev. Lett.}\ }\textbf
  {\bibinfo {volume} {77}},\ \bibinfo {pages} {3865} (\bibinfo {year}
  {1996})}\BibitemShut {NoStop}%
\bibitem [{Note1()}]{Note1}%
  \BibitemOpen
  \bibinfo {note} {Y. Xiao (private communication).}\BibitemShut {Stop}%
\bibitem [{\citenamefont {Ma}\ \emph {et~al.}(2017)\citenamefont {Ma},
  \citenamefont {Ji}, \citenamefont {Hu}, \citenamefont {Gao}, \citenamefont
  {Li}, \citenamefont {Mu},\ and\ \citenamefont {Xie}}]{Ma17SST}%
  \BibitemOpen
  \bibfield  {author} {\bibinfo {author} {\bibfnamefont {Y.}~\bibnamefont
  {Ma}}, \bibinfo {author} {\bibfnamefont {Q.}~\bibnamefont {Ji}}, \bibinfo
  {author} {\bibfnamefont {K.}~\bibnamefont {Hu}}, \bibinfo {author}
  {\bibfnamefont {B.}~\bibnamefont {Gao}}, \bibinfo {author} {\bibfnamefont
  {W.}~\bibnamefont {Li}}, \bibinfo {author} {\bibfnamefont {G.}~\bibnamefont
  {Mu}}, \ and\ \bibinfo {author} {\bibfnamefont {X.}~\bibnamefont {Xie}},\
  }\bibfield  {title} {\enquote {\bibinfo {title} {{Strong Anisotropy Effect in
  an Iron-Based Superconductor CaFe$_{0.882}$Co$_{0.118}$AsF}},}\ }\href
  {http://stacks.iop.org/0953-2048/30/i=7/a=074003} {\bibfield  {journal}
  {\bibinfo  {journal} {Supercond. Sci. Technol.}\ }\textbf {\bibinfo {volume}
  {30}},\ \bibinfo {pages} {074003} (\bibinfo {year} {2017})}\BibitemShut
  {NoStop}%
\bibitem [{\citenamefont {Richards}(1973)}]{Richards73PRB}%
  \BibitemOpen
  \bibfield  {author} {\bibinfo {author} {\bibfnamefont {F.~E.}\ \bibnamefont
  {Richards}},\ }\bibfield  {title} {\enquote {\bibinfo {title} {{Investigation
  of the Magnetoresistance Quantum Oscillations in Magnesium}},}\ }\href
  {\doibase 10.1103/PhysRevB.8.2552} {\bibfield  {journal} {\bibinfo  {journal}
  {Phys. Rev. B}\ }\textbf {\bibinfo {volume} {8}},\ \bibinfo {pages} {2552}
  (\bibinfo {year} {1973})}\BibitemShut {NoStop}%
\bibitem [{\citenamefont {Shoenberg}(1984)}]{Shoenberg84}%
  \BibitemOpen
  \bibfield  {author} {\bibinfo {author} {\bibfnamefont {D.}~\bibnamefont
  {Shoenberg}},\ }\href@noop {} {\emph {\bibinfo {title} {{Magnetic
  Oscillations in Metals}}}}\ (\bibinfo  {publisher} {Cambridge University
  Press},\ \bibinfo {address} {Cambridge},\ \bibinfo {year} {1984})\BibitemShut
  {NoStop}%
\bibitem [{\citenamefont {Mikitik}\ and\ \citenamefont
  {Sharlai}(1999)}]{Mikitik99PRL}%
  \BibitemOpen
  \bibfield  {author} {\bibinfo {author} {\bibfnamefont {G.~P.}\ \bibnamefont
  {Mikitik}}\ and\ \bibinfo {author} {\bibfnamefont {Y.~V.}\ \bibnamefont
  {Sharlai}},\ }\bibfield  {title} {\enquote {\bibinfo {title} {{Manifestation
  of Berry's Phase in Metal Physics}},}\ }\href {\doibase
  10.1103/PhysRevLett.82.2147} {\bibfield  {journal} {\bibinfo  {journal}
  {Phys. Rev. Lett.}\ }\textbf {\bibinfo {volume} {82}},\ \bibinfo {pages}
  {2147} (\bibinfo {year} {1999})}\BibitemShut {NoStop}%
\bibitem [{\citenamefont {Mikitik}\ and\ \citenamefont
  {Sharlai}(1998)}]{Mikitik98JETP}%
  \BibitemOpen
  \bibfield  {author} {\bibinfo {author} {\bibfnamefont {G.~P.}\ \bibnamefont
  {Mikitik}}\ and\ \bibinfo {author} {\bibfnamefont {Y.~V.}\ \bibnamefont
  {Sharlai}},\ }\bibfield  {title} {\enquote {\bibinfo {title} {{Semiclassical
  Energy Levels of Electrons in Metals with Band Degeneracy Lines}},}\ }\href
  {\doibase 10.1134/1.558717} {\bibfield  {journal} {\bibinfo  {journal}
  {Soviet Physics--JETP}\ }\textbf {\bibinfo {volume} {87}},\ \bibinfo {pages}
  {747} (\bibinfo {year} {1998})}\BibitemShut {NoStop}%
\bibitem [{\citenamefont {Cvetkovic}\ and\ \citenamefont
  {Vafek}(2013)}]{Cvetkovic13PRB}%
  \BibitemOpen
  \bibfield  {author} {\bibinfo {author} {\bibfnamefont {V.}~\bibnamefont
  {Cvetkovic}}\ and\ \bibinfo {author} {\bibfnamefont {O.}~\bibnamefont
  {Vafek}},\ }\bibfield  {title} {\enquote {\bibinfo {title} {Space Group
  Symmetry, Spin-Orbit Coupling, and the Low-Energy Effective Hamiltonian for
  Iron-Based Superconductors},}\ }\href {\doibase 10.1103/PhysRevB.88.134510}
  {\bibfield  {journal} {\bibinfo  {journal} {Phys. Rev. B}\ }\textbf {\bibinfo
  {volume} {88}},\ \bibinfo {pages} {134510} (\bibinfo {year}
  {2013})}\BibitemShut {NoStop}%
\bibitem [{\citenamefont {Yin}\ \emph {et~al.}(2008)\citenamefont {Yin},
  \citenamefont {Leb\`egue}, \citenamefont {Han}, \citenamefont {Neal},
  \citenamefont {Savrasov},\ and\ \citenamefont {Pickett}}]{Yin08PRL}%
  \BibitemOpen
  \bibfield  {author} {\bibinfo {author} {\bibfnamefont {Z.~P.}\ \bibnamefont
  {Yin}}, \bibinfo {author} {\bibfnamefont {S.}~\bibnamefont {Leb\`egue}},
  \bibinfo {author} {\bibfnamefont {M.~J.}\ \bibnamefont {Han}}, \bibinfo
  {author} {\bibfnamefont {B.~P.}\ \bibnamefont {Neal}}, \bibinfo {author}
  {\bibfnamefont {S.~Y.}\ \bibnamefont {Savrasov}}, \ and\ \bibinfo {author}
  {\bibfnamefont {W.~E.}\ \bibnamefont {Pickett}},\ }\bibfield  {title}
  {\enquote {\bibinfo {title} {{Electron-Hole Symmetry and Magnetic Coupling in
  Antiferromagnetic LaFeAsO}},}\ }\href {\doibase
  10.1103/PhysRevLett.101.047001} {\bibfield  {journal} {\bibinfo  {journal}
  {Phys. Rev. Lett.}\ }\textbf {\bibinfo {volume} {101}},\ \bibinfo {pages}
  {047001} (\bibinfo {year} {2008})}\BibitemShut {NoStop}%
\bibitem [{\citenamefont {Johannes}\ and\ \citenamefont
  {Mazin}(2009)}]{Johannes09PRB}%
  \BibitemOpen
  \bibfield  {author} {\bibinfo {author} {\bibfnamefont {M.~D.}\ \bibnamefont
  {Johannes}}\ and\ \bibinfo {author} {\bibfnamefont {I.~I.}\ \bibnamefont
  {Mazin}},\ }\bibfield  {title} {\enquote {\bibinfo {title} {{Microscopic
  Origin of Magnetism and Magnetic Interactions in Ferropnictides}},}\ }\href
  {\doibase 10.1103/PhysRevB.79.220510} {\bibfield  {journal} {\bibinfo
  {journal} {Phys. Rev. B}\ }\textbf {\bibinfo {volume} {79}},\ \bibinfo
  {pages} {220510} (\bibinfo {year} {2009})}\BibitemShut {NoStop}%
\bibitem [{\citenamefont {Shimojima}\ \emph {et~al.}(2010)\citenamefont
  {Shimojima}, \citenamefont {Ishizaka}, \citenamefont {Ishida}, \citenamefont
  {Katayama}, \citenamefont {Ohgushi}, \citenamefont {Kiss}, \citenamefont
  {Okawa}, \citenamefont {Togashi}, \citenamefont {Wang}, \citenamefont {Chen},
  \citenamefont {Watanabe}, \citenamefont {Kadota}, \citenamefont {Oguchi},
  \citenamefont {Chainani},\ and\ \citenamefont {Shin}}]{Shimojima10PRL}%
  \BibitemOpen
  \bibfield  {author} {\bibinfo {author} {\bibfnamefont {T.}~\bibnamefont
  {Shimojima}}, \bibinfo {author} {\bibfnamefont {K.}~\bibnamefont {Ishizaka}},
  \bibinfo {author} {\bibfnamefont {Y.}~\bibnamefont {Ishida}}, \bibinfo
  {author} {\bibfnamefont {N.}~\bibnamefont {Katayama}}, \bibinfo {author}
  {\bibfnamefont {K.}~\bibnamefont {Ohgushi}}, \bibinfo {author} {\bibfnamefont
  {T.}~\bibnamefont {Kiss}}, \bibinfo {author} {\bibfnamefont {M.}~\bibnamefont
  {Okawa}}, \bibinfo {author} {\bibfnamefont {T.}~\bibnamefont {Togashi}},
  \bibinfo {author} {\bibfnamefont {X.-Y.}\ \bibnamefont {Wang}}, \bibinfo
  {author} {\bibfnamefont {C.-T.}\ \bibnamefont {Chen}}, \bibinfo {author}
  {\bibfnamefont {S.}~\bibnamefont {Watanabe}}, \bibinfo {author}
  {\bibfnamefont {R.}~\bibnamefont {Kadota}}, \bibinfo {author} {\bibfnamefont
  {T.}~\bibnamefont {Oguchi}}, \bibinfo {author} {\bibfnamefont
  {A.}~\bibnamefont {Chainani}}, \ and\ \bibinfo {author} {\bibfnamefont
  {S.}~\bibnamefont {Shin}},\ }\bibfield  {title} {\enquote {\bibinfo {title}
  {{Orbital-Dependent Modifications of Electronic Structure across the
  Magnetostructural Transition in BaFe$_{2}$As$_{2}$}},}\ }\href {\doibase
  10.1103/PhysRevLett.104.057002} {\bibfield  {journal} {\bibinfo  {journal}
  {Phys. Rev. Lett.}\ }\textbf {\bibinfo {volume} {104}},\ \bibinfo {pages}
  {057002} (\bibinfo {year} {2010})}\BibitemShut {NoStop}%
\bibitem [{\citenamefont {Terashima}\ \emph {et~al.}(2011)\citenamefont
  {Terashima}, \citenamefont {Kurita}, \citenamefont {Tomita}, \citenamefont
  {Kihou}, \citenamefont {Lee}, \citenamefont {Tomioka}, \citenamefont {Ito},
  \citenamefont {Iyo}, \citenamefont {Eisaki}, \citenamefont {Liang},
  \citenamefont {Nakajima}, \citenamefont {Ishida}, \citenamefont {Uchida},
  \citenamefont {Harima},\ and\ \citenamefont {Uji}}]{Terashima11PRL}%
  \BibitemOpen
  \bibfield  {author} {\bibinfo {author} {\bibfnamefont {T.}~\bibnamefont
  {Terashima}}, \bibinfo {author} {\bibfnamefont {N.}~\bibnamefont {Kurita}},
  \bibinfo {author} {\bibfnamefont {M.}~\bibnamefont {Tomita}}, \bibinfo
  {author} {\bibfnamefont {K.}~\bibnamefont {Kihou}}, \bibinfo {author}
  {\bibfnamefont {C.~H.}\ \bibnamefont {Lee}}, \bibinfo {author} {\bibfnamefont
  {Y.}~\bibnamefont {Tomioka}}, \bibinfo {author} {\bibfnamefont
  {T.}~\bibnamefont {Ito}}, \bibinfo {author} {\bibfnamefont {A.}~\bibnamefont
  {Iyo}}, \bibinfo {author} {\bibfnamefont {H.}~\bibnamefont {Eisaki}},
  \bibinfo {author} {\bibfnamefont {T.}~\bibnamefont {Liang}}, \bibinfo
  {author} {\bibfnamefont {M.}~\bibnamefont {Nakajima}}, \bibinfo {author}
  {\bibfnamefont {S.}~\bibnamefont {Ishida}}, \bibinfo {author} {\bibfnamefont
  {S.-i.}\ \bibnamefont {Uchida}}, \bibinfo {author} {\bibfnamefont
  {H.}~\bibnamefont {Harima}}, \ and\ \bibinfo {author} {\bibfnamefont
  {S.}~\bibnamefont {Uji}},\ }\bibfield  {title} {\enquote {\bibinfo {title}
  {{Complete Fermi Surface in ${\mathrm{BaFe}}_{2}{\mathrm{As}}_{2}$ Observed
  via Shubnikov--de Haas Oscillation Measurements on Detwinned Single
  Crystals}},}\ }\href {\doibase 10.1103/PhysRevLett.107.176402} {\bibfield
  {journal} {\bibinfo  {journal} {Phys. Rev. Lett.}\ }\textbf {\bibinfo
  {volume} {107}},\ \bibinfo {pages} {176402} (\bibinfo {year}
  {2011})}\BibitemShut {NoStop}%
\bibitem [{\citenamefont {Terashima}\ \emph {et~al.}(2014)\citenamefont
  {Terashima}, \citenamefont {Kikugawa}, \citenamefont {Kiswandhi},
  \citenamefont {Choi}, \citenamefont {Brooks}, \citenamefont {Kasahara},
  \citenamefont {Watashige}, \citenamefont {Ikeda}, \citenamefont {Shibauchi},
  \citenamefont {Matsuda}, \citenamefont {Wolf}, \citenamefont {B\"ohmer},
  \citenamefont {Hardy}, \citenamefont {Meingast}, \citenamefont {L\"ohneysen},
  \citenamefont {Suzuki}, \citenamefont {Arita},\ and\ \citenamefont
  {Uji}}]{Terashima14PRB}%
  \BibitemOpen
  \bibfield  {author} {\bibinfo {author} {\bibfnamefont {T.}~\bibnamefont
  {Terashima}}, \bibinfo {author} {\bibfnamefont {N.}~\bibnamefont {Kikugawa}},
  \bibinfo {author} {\bibfnamefont {A.}~\bibnamefont {Kiswandhi}}, \bibinfo
  {author} {\bibfnamefont {E.-S.}\ \bibnamefont {Choi}}, \bibinfo {author}
  {\bibfnamefont {J.~S.}\ \bibnamefont {Brooks}}, \bibinfo {author}
  {\bibfnamefont {S.}~\bibnamefont {Kasahara}}, \bibinfo {author}
  {\bibfnamefont {T.}~\bibnamefont {Watashige}}, \bibinfo {author}
  {\bibfnamefont {H.}~\bibnamefont {Ikeda}}, \bibinfo {author} {\bibfnamefont
  {T.}~\bibnamefont {Shibauchi}}, \bibinfo {author} {\bibfnamefont
  {Y.}~\bibnamefont {Matsuda}}, \bibinfo {author} {\bibfnamefont
  {T.}~\bibnamefont {Wolf}}, \bibinfo {author} {\bibfnamefont {A.~E.}\
  \bibnamefont {B\"ohmer}}, \bibinfo {author} {\bibfnamefont {F.}~\bibnamefont
  {Hardy}}, \bibinfo {author} {\bibfnamefont {C.}~\bibnamefont {Meingast}},
  \bibinfo {author} {\bibfnamefont {H.~v.}\ \bibnamefont {L\"ohneysen}},
  \bibinfo {author} {\bibfnamefont {M.-T.}\ \bibnamefont {Suzuki}}, \bibinfo
  {author} {\bibfnamefont {R.}~\bibnamefont {Arita}}, \ and\ \bibinfo {author}
  {\bibfnamefont {S.}~\bibnamefont {Uji}},\ }\bibfield  {title} {\enquote
  {\bibinfo {title} {{Anomalous Fermi Surface in FeSe Seen by Shubnikov--de
  Haas Oscillation Measurements}},}\ }\href {\doibase
  10.1103/PhysRevB.90.144517} {\bibfield  {journal} {\bibinfo  {journal} {Phys.
  Rev. B}\ }\textbf {\bibinfo {volume} {90}},\ \bibinfo {pages} {144517}
  (\bibinfo {year} {2014})}\BibitemShut {NoStop}%
\bibitem [{\citenamefont {Tan}\ \emph {et~al.}(2013)\citenamefont {Tan},
  \citenamefont {Zhang}, \citenamefont {Xia}, \citenamefont {Ye}, ,
  \citenamefont {Chen}, \citenamefont {Xie}, \citenamefont {Peng},
  \citenamefont {Xu}, \citenamefont {Qin~Fan}, \citenamefont {Jiang},
  \citenamefont {Zhang}, \citenamefont {Lai}, \citenamefont {Xiang},
  \citenamefont {Hu}, \citenamefont {Xie},\ and\ \citenamefont
  {Feng}}]{Tan13NatMat}%
  \BibitemOpen
  \bibfield  {author} {\bibinfo {author} {\bibfnamefont {S.}~\bibnamefont
  {Tan}}, \bibinfo {author} {\bibfnamefont {Y.}~\bibnamefont {Zhang}}, \bibinfo
  {author} {\bibfnamefont {M.}~\bibnamefont {Xia}}, \bibinfo {author}
  {\bibfnamefont {Z.}~\bibnamefont {Ye}}, , \bibinfo {author} {\bibfnamefont
  {F.}~\bibnamefont {Chen}}, \bibinfo {author} {\bibfnamefont {X.}~\bibnamefont
  {Xie}}, \bibinfo {author} {\bibfnamefont {R.}~\bibnamefont {Peng}}, \bibinfo
  {author} {\bibfnamefont {D.}~\bibnamefont {Xu}}, \bibinfo {author}
  {\bibfnamefont {H.~X.}\ \bibnamefont {Qin~Fan}}, \bibinfo {author}
  {\bibfnamefont {J.}~\bibnamefont {Jiang}}, \bibinfo {author} {\bibfnamefont
  {T.}~\bibnamefont {Zhang}}, \bibinfo {author} {\bibfnamefont
  {X.}~\bibnamefont {Lai}}, \bibinfo {author} {\bibfnamefont {T.}~\bibnamefont
  {Xiang}}, \bibinfo {author} {\bibfnamefont {J.}~\bibnamefont {Hu}}, \bibinfo
  {author} {\bibfnamefont {B.}~\bibnamefont {Xie}}, \ and\ \bibinfo {author}
  {\bibfnamefont {D.}~\bibnamefont {Feng}},\ }\bibfield  {title} {\enquote
  {\bibinfo {title} {{Interface-Induced Superconductivity and Strain-Dependent
  Spin Density Waves in FeSe/SrTiO$_3$ Thin Films}},}\ }\href@noop {}
  {\bibfield  {journal} {\bibinfo  {journal} {Nat. Mater.}\ }\textbf {\bibinfo
  {volume} {12}},\ \bibinfo {pages} {634} (\bibinfo {year} {2013})}\BibitemShut
  {NoStop}%
\bibitem [{\citenamefont {Maletz}\ \emph {et~al.}(2014)\citenamefont {Maletz},
  \citenamefont {Zabolotnyy}, \citenamefont {Evtushinsky}, \citenamefont
  {Thirupathaiah}, \citenamefont {Wolter}, \citenamefont {Harnagea},
  \citenamefont {Yaresko}, \citenamefont {Vasiliev}, \citenamefont {Chareev},
  \citenamefont {B\"ohmer}, \citenamefont {Hardy}, \citenamefont {Wolf},
  \citenamefont {Meingast}, \citenamefont {Rienks}, \citenamefont {B\"uchner},\
  and\ \citenamefont {Borisenko}}]{Maletz14PRB}%
  \BibitemOpen
  \bibfield  {author} {\bibinfo {author} {\bibfnamefont {J.}~\bibnamefont
  {Maletz}}, \bibinfo {author} {\bibfnamefont {V.~B.}\ \bibnamefont
  {Zabolotnyy}}, \bibinfo {author} {\bibfnamefont {D.~V.}\ \bibnamefont
  {Evtushinsky}}, \bibinfo {author} {\bibfnamefont {S.}~\bibnamefont
  {Thirupathaiah}}, \bibinfo {author} {\bibfnamefont {A.~U.~B.}\ \bibnamefont
  {Wolter}}, \bibinfo {author} {\bibfnamefont {L.}~\bibnamefont {Harnagea}},
  \bibinfo {author} {\bibfnamefont {A.~N.}\ \bibnamefont {Yaresko}}, \bibinfo
  {author} {\bibfnamefont {A.~N.}\ \bibnamefont {Vasiliev}}, \bibinfo {author}
  {\bibfnamefont {D.~A.}\ \bibnamefont {Chareev}}, \bibinfo {author}
  {\bibfnamefont {A.~E.}\ \bibnamefont {B\"ohmer}}, \bibinfo {author}
  {\bibfnamefont {F.}~\bibnamefont {Hardy}}, \bibinfo {author} {\bibfnamefont
  {T.}~\bibnamefont {Wolf}}, \bibinfo {author} {\bibfnamefont {C.}~\bibnamefont
  {Meingast}}, \bibinfo {author} {\bibfnamefont {E.~D.~L.}\ \bibnamefont
  {Rienks}}, \bibinfo {author} {\bibfnamefont {B.}~\bibnamefont {B\"uchner}}, \
  and\ \bibinfo {author} {\bibfnamefont {S.~V.}\ \bibnamefont {Borisenko}},\
  }\bibfield  {title} {\enquote {\bibinfo {title} {{Unusual Band
  Renormalization in the Simplest Iron-Based Superconductor
  ${\text{FeSe}}_{1\ensuremath{-}x}$}},}\ }\href {\doibase
  10.1103/PhysRevB.89.220506} {\bibfield  {journal} {\bibinfo  {journal} {Phys.
  Rev. B}\ }\textbf {\bibinfo {volume} {89}},\ \bibinfo {pages} {220506}
  (\bibinfo {year} {2014})}\BibitemShut {NoStop}%
\bibitem [{\citenamefont {Nakayama}\ \emph {et~al.}(2014)\citenamefont
  {Nakayama}, \citenamefont {Miyata}, \citenamefont {Phan}, \citenamefont
  {Sato}, \citenamefont {Tanabe}, \citenamefont {Urata}, \citenamefont
  {Tanigaki},\ and\ \citenamefont {Takahashi}}]{Nakayama14PRL}%
  \BibitemOpen
  \bibfield  {author} {\bibinfo {author} {\bibfnamefont {K.}~\bibnamefont
  {Nakayama}}, \bibinfo {author} {\bibfnamefont {Y.}~\bibnamefont {Miyata}},
  \bibinfo {author} {\bibfnamefont {G.~N.}\ \bibnamefont {Phan}}, \bibinfo
  {author} {\bibfnamefont {T.}~\bibnamefont {Sato}}, \bibinfo {author}
  {\bibfnamefont {Y.}~\bibnamefont {Tanabe}}, \bibinfo {author} {\bibfnamefont
  {T.}~\bibnamefont {Urata}}, \bibinfo {author} {\bibfnamefont
  {K.}~\bibnamefont {Tanigaki}}, \ and\ \bibinfo {author} {\bibfnamefont
  {T.}~\bibnamefont {Takahashi}},\ }\bibfield  {title} {\enquote {\bibinfo
  {title} {{Reconstruction of Band Structure Induced by Electronic Nematicity
  in an FeSe Superconductor}},}\ }\href {\doibase
  10.1103/PhysRevLett.113.237001} {\bibfield  {journal} {\bibinfo  {journal}
  {Phys. Rev. Lett.}\ }\textbf {\bibinfo {volume} {113}},\ \bibinfo {pages}
  {237001} (\bibinfo {year} {2014})}\BibitemShut {NoStop}%
\bibitem [{\citenamefont {Shimojima}\ \emph {et~al.}(2014)\citenamefont
  {Shimojima}, \citenamefont {Suzuki}, \citenamefont {Sonobe}, \citenamefont
  {Nakamura}, \citenamefont {Sakano}, \citenamefont {Omachi}, \citenamefont
  {Yoshioka}, \citenamefont {Kuwata-Gonokami}, \citenamefont {Ono},
  \citenamefont {Kumigashira}, \citenamefont {B\"ohmer}, \citenamefont {Hardy},
  \citenamefont {Wolf}, \citenamefont {Meingast}, \citenamefont {L\"ohneysen},
  \citenamefont {Ikeda},\ and\ \citenamefont {Ishizaka}}]{Shimojima14PRB}%
  \BibitemOpen
  \bibfield  {author} {\bibinfo {author} {\bibfnamefont {T.}~\bibnamefont
  {Shimojima}}, \bibinfo {author} {\bibfnamefont {Y.}~\bibnamefont {Suzuki}},
  \bibinfo {author} {\bibfnamefont {T.}~\bibnamefont {Sonobe}}, \bibinfo
  {author} {\bibfnamefont {A.}~\bibnamefont {Nakamura}}, \bibinfo {author}
  {\bibfnamefont {M.}~\bibnamefont {Sakano}}, \bibinfo {author} {\bibfnamefont
  {J.}~\bibnamefont {Omachi}}, \bibinfo {author} {\bibfnamefont
  {K.}~\bibnamefont {Yoshioka}}, \bibinfo {author} {\bibfnamefont
  {M.}~\bibnamefont {Kuwata-Gonokami}}, \bibinfo {author} {\bibfnamefont
  {K.}~\bibnamefont {Ono}}, \bibinfo {author} {\bibfnamefont {H.}~\bibnamefont
  {Kumigashira}}, \bibinfo {author} {\bibfnamefont {A.~E.}\ \bibnamefont
  {B\"ohmer}}, \bibinfo {author} {\bibfnamefont {F.}~\bibnamefont {Hardy}},
  \bibinfo {author} {\bibfnamefont {T.}~\bibnamefont {Wolf}}, \bibinfo {author}
  {\bibfnamefont {C.}~\bibnamefont {Meingast}}, \bibinfo {author}
  {\bibfnamefont {H.~v.}\ \bibnamefont {L\"ohneysen}}, \bibinfo {author}
  {\bibfnamefont {H.}~\bibnamefont {Ikeda}}, \ and\ \bibinfo {author}
  {\bibfnamefont {K.}~\bibnamefont {Ishizaka}},\ }\bibfield  {title} {\enquote
  {\bibinfo {title} {{Lifting of \textit{xz}/\textit{yz} Orbital Degeneracy at
  the Structural Transition in Detwinned FeSe}},}\ }\href {\doibase
  10.1103/PhysRevB.90.121111} {\bibfield  {journal} {\bibinfo  {journal} {Phys.
  Rev. B}\ }\textbf {\bibinfo {volume} {90}},\ \bibinfo {pages} {121111}
  (\bibinfo {year} {2014})}\BibitemShut {NoStop}%
\bibitem [{\citenamefont {Audouard}\ \emph {et~al.}(2015)\citenamefont
  {Audouard}, \citenamefont {Duc}, \citenamefont {Drigo}, \citenamefont
  {Toulemonde}, \citenamefont {Karlsson}, \citenamefont {Strobel},\ and\
  \citenamefont {Sulpice}}]{Audouard15EPL}%
  \BibitemOpen
  \bibfield  {author} {\bibinfo {author} {\bibfnamefont {A.}~\bibnamefont
  {Audouard}}, \bibinfo {author} {\bibfnamefont {F.}~\bibnamefont {Duc}},
  \bibinfo {author} {\bibfnamefont {L.}~\bibnamefont {Drigo}}, \bibinfo
  {author} {\bibfnamefont {P.}~\bibnamefont {Toulemonde}}, \bibinfo {author}
  {\bibfnamefont {S.}~\bibnamefont {Karlsson}}, \bibinfo {author}
  {\bibfnamefont {P.}~\bibnamefont {Strobel}}, \ and\ \bibinfo {author}
  {\bibfnamefont {A.}~\bibnamefont {Sulpice}},\ }\bibfield  {title} {\enquote
  {\bibinfo {title} {{Quantum Oscillations and Upper Critical Magnetic Field of
  the Iron-Based Superconductor FeSe}},}\ }\href
  {http://stacks.iop.org/0295-5075/109/i=2/a=27003} {\bibfield  {journal}
  {\bibinfo  {journal} {EPL}\ }\textbf {\bibinfo {volume} {109}},\ \bibinfo
  {pages} {27003} (\bibinfo {year} {2015})}\BibitemShut {NoStop}%
\bibitem [{\citenamefont {Watson}\ \emph
  {et~al.}(2015{\natexlab{a}})\citenamefont {Watson}, \citenamefont {Kim},
  \citenamefont {Haghighirad}, \citenamefont {Davies}, \citenamefont
  {McCollam}, \citenamefont {Narayanan}, \citenamefont {Blake}, \citenamefont
  {Chen}, \citenamefont {Ghannadzadeh}, \citenamefont {Schofield},
  \citenamefont {Hoesch}, \citenamefont {Meingast}, \citenamefont {Wolf},\ and\
  \citenamefont {Coldea}}]{Watson15PRB}%
  \BibitemOpen
  \bibfield  {author} {\bibinfo {author} {\bibfnamefont {M.~D.}\ \bibnamefont
  {Watson}}, \bibinfo {author} {\bibfnamefont {T.~K.}\ \bibnamefont {Kim}},
  \bibinfo {author} {\bibfnamefont {A.~A.}\ \bibnamefont {Haghighirad}},
  \bibinfo {author} {\bibfnamefont {N.~R.}\ \bibnamefont {Davies}}, \bibinfo
  {author} {\bibfnamefont {A.}~\bibnamefont {McCollam}}, \bibinfo {author}
  {\bibfnamefont {A.}~\bibnamefont {Narayanan}}, \bibinfo {author}
  {\bibfnamefont {S.~F.}\ \bibnamefont {Blake}}, \bibinfo {author}
  {\bibfnamefont {Y.~L.}\ \bibnamefont {Chen}}, \bibinfo {author}
  {\bibfnamefont {S.}~\bibnamefont {Ghannadzadeh}}, \bibinfo {author}
  {\bibfnamefont {A.~J.}\ \bibnamefont {Schofield}}, \bibinfo {author}
  {\bibfnamefont {M.}~\bibnamefont {Hoesch}}, \bibinfo {author} {\bibfnamefont
  {C.}~\bibnamefont {Meingast}}, \bibinfo {author} {\bibfnamefont
  {T.}~\bibnamefont {Wolf}}, \ and\ \bibinfo {author} {\bibfnamefont {A.~I.}\
  \bibnamefont {Coldea}},\ }\bibfield  {title} {\enquote {\bibinfo {title}
  {{Emergence of the Nematic Electronic State in FeSe}},}\ }\href {\doibase
  10.1103/PhysRevB.91.155106} {\bibfield  {journal} {\bibinfo  {journal} {Phys.
  Rev. B}\ }\textbf {\bibinfo {volume} {91}},\ \bibinfo {pages} {155106}
  (\bibinfo {year} {2015}{\natexlab{a}})}\BibitemShut {NoStop}%
\bibitem [{\citenamefont {Watson}\ \emph
  {et~al.}(2015{\natexlab{b}})\citenamefont {Watson}, \citenamefont
  {Yamashita}, \citenamefont {Kasahara}, \citenamefont {Knafo}, \citenamefont
  {Nardone}, \citenamefont {B\'eard}, \citenamefont {Hardy}, \citenamefont
  {McCollam}, \citenamefont {Narayanan}, \citenamefont {Blake}, \citenamefont
  {Wolf}, \citenamefont {Haghighirad}, \citenamefont {Meingast}, \citenamefont
  {Schofield}, \citenamefont {v.~L\"ohneysen}, \citenamefont {Matsuda},
  \citenamefont {Coldea},\ and\ \citenamefont {Shibauchi}}]{Watson15PRL}%
  \BibitemOpen
  \bibfield  {author} {\bibinfo {author} {\bibfnamefont {M.~D.}\ \bibnamefont
  {Watson}}, \bibinfo {author} {\bibfnamefont {T.}~\bibnamefont {Yamashita}},
  \bibinfo {author} {\bibfnamefont {S.}~\bibnamefont {Kasahara}}, \bibinfo
  {author} {\bibfnamefont {W.}~\bibnamefont {Knafo}}, \bibinfo {author}
  {\bibfnamefont {M.}~\bibnamefont {Nardone}}, \bibinfo {author} {\bibfnamefont
  {J.}~\bibnamefont {B\'eard}}, \bibinfo {author} {\bibfnamefont
  {F.}~\bibnamefont {Hardy}}, \bibinfo {author} {\bibfnamefont
  {A.}~\bibnamefont {McCollam}}, \bibinfo {author} {\bibfnamefont
  {A.}~\bibnamefont {Narayanan}}, \bibinfo {author} {\bibfnamefont {S.~F.}\
  \bibnamefont {Blake}}, \bibinfo {author} {\bibfnamefont {T.}~\bibnamefont
  {Wolf}}, \bibinfo {author} {\bibfnamefont {A.~A.}\ \bibnamefont
  {Haghighirad}}, \bibinfo {author} {\bibfnamefont {C.}~\bibnamefont
  {Meingast}}, \bibinfo {author} {\bibfnamefont {A.~J.}\ \bibnamefont
  {Schofield}}, \bibinfo {author} {\bibfnamefont {H.}~\bibnamefont
  {v.~L\"ohneysen}}, \bibinfo {author} {\bibfnamefont {Y.}~\bibnamefont
  {Matsuda}}, \bibinfo {author} {\bibfnamefont {A.~I.}\ \bibnamefont {Coldea}},
  \ and\ \bibinfo {author} {\bibfnamefont {T.}~\bibnamefont {Shibauchi}},\
  }\bibfield  {title} {\enquote {\bibinfo {title} {{Dichotomy between the Hole
  and Electron Behavior in Multiband Superconductor FeSe Probed by Ultrahigh
  Magnetic Fields}},}\ }\href {\doibase 10.1103/PhysRevLett.115.027006}
  {\bibfield  {journal} {\bibinfo  {journal} {Phys. Rev. Lett.}\ }\textbf
  {\bibinfo {volume} {115}},\ \bibinfo {pages} {027006} (\bibinfo {year}
  {2015}{\natexlab{b}})}\BibitemShut {NoStop}%
\bibitem [{\citenamefont {McQueen}\ \emph {et~al.}(2009)\citenamefont
  {McQueen}, \citenamefont {Williams}, \citenamefont {Stephens}, \citenamefont
  {Tao}, \citenamefont {Zhu}, \citenamefont {Ksenofontov}, \citenamefont
  {Casper}, \citenamefont {Felser},\ and\ \citenamefont {Cava}}]{McQueen09PRL}%
  \BibitemOpen
  \bibfield  {author} {\bibinfo {author} {\bibfnamefont {T.~M.}\ \bibnamefont
  {McQueen}}, \bibinfo {author} {\bibfnamefont {A.~J.}\ \bibnamefont
  {Williams}}, \bibinfo {author} {\bibfnamefont {P.~W.}\ \bibnamefont
  {Stephens}}, \bibinfo {author} {\bibfnamefont {J.}~\bibnamefont {Tao}},
  \bibinfo {author} {\bibfnamefont {Y.}~\bibnamefont {Zhu}}, \bibinfo {author}
  {\bibfnamefont {V.}~\bibnamefont {Ksenofontov}}, \bibinfo {author}
  {\bibfnamefont {F.}~\bibnamefont {Casper}}, \bibinfo {author} {\bibfnamefont
  {C.}~\bibnamefont {Felser}}, \ and\ \bibinfo {author} {\bibfnamefont {R.~J.}\
  \bibnamefont {Cava}},\ }\bibfield  {title} {\enquote {\bibinfo {title}
  {{Tetragonal-to-Orthorhombic Structural Phase Transition at 90 K in the
  Superconductor ${\mathrm{Fe}}_{1.01}\mathrm{Se}$}},}\ }\href {\doibase
  10.1103/PhysRevLett.103.057002} {\bibfield  {journal} {\bibinfo  {journal}
  {Phys. Rev. Lett.}\ }\textbf {\bibinfo {volume} {103}},\ \bibinfo {pages}
  {057002} (\bibinfo {year} {2009})}\BibitemShut {NoStop}%
\bibitem [{\citenamefont {Ran}\ \emph {et~al.}(2009)\citenamefont {Ran},
  \citenamefont {Wang}, \citenamefont {Zhai}, \citenamefont {Vishwanath},\ and\
  \citenamefont {Lee}}]{Ran09PRB}%
  \BibitemOpen
  \bibfield  {author} {\bibinfo {author} {\bibfnamefont {Y.}~\bibnamefont
  {Ran}}, \bibinfo {author} {\bibfnamefont {F.}~\bibnamefont {Wang}}, \bibinfo
  {author} {\bibfnamefont {H.}~\bibnamefont {Zhai}}, \bibinfo {author}
  {\bibfnamefont {A.}~\bibnamefont {Vishwanath}}, \ and\ \bibinfo {author}
  {\bibfnamefont {D.-H.}\ \bibnamefont {Lee}},\ }\bibfield  {title} {\enquote
  {\bibinfo {title} {{Nodal Spin Density Wave and Band Topology of the
  FeAs-Based Materials}},}\ }\href {\doibase 10.1103/PhysRevB.79.014505}
  {\bibfield  {journal} {\bibinfo  {journal} {Phys. Rev. B}\ }\textbf {\bibinfo
  {volume} {79}},\ \bibinfo {pages} {014505} (\bibinfo {year}
  {2009})}\BibitemShut {NoStop}%
\bibitem [{\citenamefont {Morinari}\ \emph {et~al.}(2010)\citenamefont
  {Morinari}, \citenamefont {Kaneshita},\ and\ \citenamefont
  {Tohyama}}]{Morinari10PRL}%
  \BibitemOpen
  \bibfield  {author} {\bibinfo {author} {\bibfnamefont {T.}~\bibnamefont
  {Morinari}}, \bibinfo {author} {\bibfnamefont {E.}~\bibnamefont {Kaneshita}},
  \ and\ \bibinfo {author} {\bibfnamefont {T.}~\bibnamefont {Tohyama}},\
  }\bibfield  {title} {\enquote {\bibinfo {title} {{Topological and Transport
  Properties of Dirac Fermions in an Antiferromagnetic Metallic Phase of
  Iron-Based Superconductors}},}\ }\href {\doibase
  10.1103/PhysRevLett.105.037203} {\bibfield  {journal} {\bibinfo  {journal}
  {Phys. Rev. Lett.}\ }\textbf {\bibinfo {volume} {105}},\ \bibinfo {pages}
  {037203} (\bibinfo {year} {2010})}\BibitemShut {NoStop}%
\bibitem [{\citenamefont {Huynh}\ \emph {et~al.}(2011)\citenamefont {Huynh},
  \citenamefont {Tanabe},\ and\ \citenamefont {Tanigaki}}]{Huynh11PRL}%
  \BibitemOpen
  \bibfield  {author} {\bibinfo {author} {\bibfnamefont {K.~K.}\ \bibnamefont
  {Huynh}}, \bibinfo {author} {\bibfnamefont {Y.}~\bibnamefont {Tanabe}}, \
  and\ \bibinfo {author} {\bibfnamefont {K.}~\bibnamefont {Tanigaki}},\
  }\bibfield  {title} {\enquote {\bibinfo {title} {{Both Electron and Hole
  Dirac Cone States in Ba(FeAs)$_{2}$ Confirmed by Magnetoresistance}},}\
  }\href {\doibase 10.1103/PhysRevLett.106.217004} {\bibfield  {journal}
  {\bibinfo  {journal} {Phys. Rev. Lett.}\ }\textbf {\bibinfo {volume} {106}},\
  \bibinfo {pages} {217004} (\bibinfo {year} {2011})}\BibitemShut {NoStop}%
\bibitem [{\citenamefont {Tanabe}\ \emph {et~al.}(2011)\citenamefont {Tanabe},
  \citenamefont {Huynh}, \citenamefont {Heguri}, \citenamefont {Mu},
  \citenamefont {Urata}, \citenamefont {Xu}, \citenamefont {Nouchi},
  \citenamefont {Mitoma},\ and\ \citenamefont {Tanigaki}}]{Tanabe11PRB}%
  \BibitemOpen
  \bibfield  {author} {\bibinfo {author} {\bibfnamefont {Y.}~\bibnamefont
  {Tanabe}}, \bibinfo {author} {\bibfnamefont {K.~K.}\ \bibnamefont {Huynh}},
  \bibinfo {author} {\bibfnamefont {S.}~\bibnamefont {Heguri}}, \bibinfo
  {author} {\bibfnamefont {G.}~\bibnamefont {Mu}}, \bibinfo {author}
  {\bibfnamefont {T.}~\bibnamefont {Urata}}, \bibinfo {author} {\bibfnamefont
  {J.}~\bibnamefont {Xu}}, \bibinfo {author} {\bibfnamefont {R.}~\bibnamefont
  {Nouchi}}, \bibinfo {author} {\bibfnamefont {N.}~\bibnamefont {Mitoma}}, \
  and\ \bibinfo {author} {\bibfnamefont {K.}~\bibnamefont {Tanigaki}},\
  }\bibfield  {title} {\enquote {\bibinfo {title} {{Coexistence of Dirac-Cone
  States and Superconductivity in Iron Pnictide
  Ba(Fe${}_{1\ensuremath{-}x}$Ru${}_{x}$As)${}_{2}$}},}\ }\href {\doibase
  10.1103/PhysRevB.84.100508} {\bibfield  {journal} {\bibinfo  {journal} {Phys.
  Rev. B}\ }\textbf {\bibinfo {volume} {84}},\ \bibinfo {pages} {100508}
  (\bibinfo {year} {2011})}\BibitemShut {NoStop}%
\bibitem [{\citenamefont {Pallecchi}\ \emph {et~al.}(2011)\citenamefont
  {Pallecchi}, \citenamefont {Bernardini}, \citenamefont {Tropeano},
  \citenamefont {Palenzona}, \citenamefont {Martinelli}, \citenamefont
  {Ferdeghini}, \citenamefont {Vignolo}, \citenamefont {Massidda},\ and\
  \citenamefont {Putti}}]{Pallecchi11PRB}%
  \BibitemOpen
  \bibfield  {author} {\bibinfo {author} {\bibfnamefont {I.}~\bibnamefont
  {Pallecchi}}, \bibinfo {author} {\bibfnamefont {F.}~\bibnamefont
  {Bernardini}}, \bibinfo {author} {\bibfnamefont {M.}~\bibnamefont
  {Tropeano}}, \bibinfo {author} {\bibfnamefont {A.}~\bibnamefont {Palenzona}},
  \bibinfo {author} {\bibfnamefont {A.}~\bibnamefont {Martinelli}}, \bibinfo
  {author} {\bibfnamefont {C.}~\bibnamefont {Ferdeghini}}, \bibinfo {author}
  {\bibfnamefont {M.}~\bibnamefont {Vignolo}}, \bibinfo {author} {\bibfnamefont
  {S.}~\bibnamefont {Massidda}}, \ and\ \bibinfo {author} {\bibfnamefont
  {M.}~\bibnamefont {Putti}},\ }\bibfield  {title} {\enquote {\bibinfo {title}
  {{Magnetotransport in La(Fe,Ru)AsO as a Probe of Band Structure and
  Mobility}},}\ }\href {\doibase 10.1103/PhysRevB.84.134524} {\bibfield
  {journal} {\bibinfo  {journal} {Phys. Rev. B}\ }\textbf {\bibinfo {volume}
  {84}},\ \bibinfo {pages} {134524} (\bibinfo {year} {2011})}\BibitemShut
  {NoStop}%
\bibitem [{\citenamefont {Richard}\ \emph {et~al.}(2010)\citenamefont
  {Richard}, \citenamefont {Nakayama}, \citenamefont {Sato}, \citenamefont
  {Neupane}, \citenamefont {Xu}, \citenamefont {Bowen}, \citenamefont {Chen},
  \citenamefont {Luo}, \citenamefont {Wang}, \citenamefont {Dai}, \citenamefont
  {Fang}, \citenamefont {Ding},\ and\ \citenamefont
  {Takahashi}}]{Richard10PRL}%
  \BibitemOpen
  \bibfield  {author} {\bibinfo {author} {\bibfnamefont {P.}~\bibnamefont
  {Richard}}, \bibinfo {author} {\bibfnamefont {K.}~\bibnamefont {Nakayama}},
  \bibinfo {author} {\bibfnamefont {T.}~\bibnamefont {Sato}}, \bibinfo {author}
  {\bibfnamefont {M.}~\bibnamefont {Neupane}}, \bibinfo {author} {\bibfnamefont
  {Y.-M.}\ \bibnamefont {Xu}}, \bibinfo {author} {\bibfnamefont {J.~H.}\
  \bibnamefont {Bowen}}, \bibinfo {author} {\bibfnamefont {G.~F.}\ \bibnamefont
  {Chen}}, \bibinfo {author} {\bibfnamefont {J.~L.}\ \bibnamefont {Luo}},
  \bibinfo {author} {\bibfnamefont {N.~L.}\ \bibnamefont {Wang}}, \bibinfo
  {author} {\bibfnamefont {X.}~\bibnamefont {Dai}}, \bibinfo {author}
  {\bibfnamefont {Z.}~\bibnamefont {Fang}}, \bibinfo {author} {\bibfnamefont
  {H.}~\bibnamefont {Ding}}, \ and\ \bibinfo {author} {\bibfnamefont
  {T.}~\bibnamefont {Takahashi}},\ }\bibfield  {title} {\enquote {\bibinfo
  {title} {{Observation of Dirac Cone Electronic Dispersion in
  BaFe$_{2}$As$_{2}$}},}\ }\href {\doibase 10.1103/PhysRevLett.104.137001}
  {\bibfield  {journal} {\bibinfo  {journal} {Phys. Rev. Lett.}\ }\textbf
  {\bibinfo {volume} {104}},\ \bibinfo {pages} {137001} (\bibinfo {year}
  {2010})}\BibitemShut {NoStop}%
\bibitem [{\citenamefont {Yi}\ \emph {et~al.}(2011)\citenamefont {Yi},
  \citenamefont {Lu}, \citenamefont {Chu}, \citenamefont {Analytis},
  \citenamefont {Sorini}, \citenamefont {Kemper}, \citenamefont {Mo},
  \citenamefont {Moore}, \citenamefont {Hashimoto}, \citenamefont {Lee},
  \citenamefont {Hussain}, \citenamefont {Devereaux}, \citenamefont {Fisher},\
  and\ \citenamefont {Shen}}]{Yi11PNAS}%
  \BibitemOpen
  \bibfield  {author} {\bibinfo {author} {\bibfnamefont {M.}~\bibnamefont
  {Yi}}, \bibinfo {author} {\bibfnamefont {D.~H.}\ \bibnamefont {Lu}}, \bibinfo
  {author} {\bibfnamefont {J.-H.}\ \bibnamefont {Chu}}, \bibinfo {author}
  {\bibfnamefont {J.~G.}\ \bibnamefont {Analytis}}, \bibinfo {author}
  {\bibfnamefont {A.~P.}\ \bibnamefont {Sorini}}, \bibinfo {author}
  {\bibfnamefont {A.~F.}\ \bibnamefont {Kemper}}, \bibinfo {author}
  {\bibfnamefont {S.-K.}\ \bibnamefont {Mo}}, \bibinfo {author} {\bibfnamefont
  {R.~G.}\ \bibnamefont {Moore}}, \bibinfo {author} {\bibfnamefont
  {M.}~\bibnamefont {Hashimoto}}, \bibinfo {author} {\bibfnamefont {W.~S.}\
  \bibnamefont {Lee}}, \bibinfo {author} {\bibfnamefont {Z.}~\bibnamefont
  {Hussain}}, \bibinfo {author} {\bibfnamefont {T.~P.}\ \bibnamefont
  {Devereaux}}, \bibinfo {author} {\bibfnamefont {I.~R.}\ \bibnamefont
  {Fisher}}, \ and\ \bibinfo {author} {\bibfnamefont {Z.-X.}\ \bibnamefont
  {Shen}},\ }\bibfield  {title} {\enquote {\bibinfo {title} {{Symmetry-Breaking
  Orbital Anisotropy Observed for Detwinned Ba(Fe$_{1-x}$Co$_x$)$_2$As$_2$
  above the Spin Density Wave Transition}},}\ }\href@noop {} {\bibfield
  {journal} {\bibinfo  {journal} {Proc. Nat. Acad. Sci. U. S. A.}\ }\textbf
  {\bibinfo {volume} {108}},\ \bibinfo {pages} {6878} (\bibinfo {year}
  {2011})}\BibitemShut {NoStop}%
\bibitem [{\citenamefont {Nakajima}\ \emph {et~al.}(2011)\citenamefont
  {Nakajima}, \citenamefont {Liang}, \citenamefont {Ishida}, \citenamefont
  {Tomioka}, \citenamefont {Kihou}, \citenamefont {Lee}, \citenamefont {Iyo},
  \citenamefont {Eisaki}, \citenamefont {Kakeshita}, \citenamefont {Ito},\ and\
  \citenamefont {Uchida}}]{Nakajima11PNAS}%
  \BibitemOpen
  \bibfield  {author} {\bibinfo {author} {\bibfnamefont {M.}~\bibnamefont
  {Nakajima}}, \bibinfo {author} {\bibfnamefont {T.}~\bibnamefont {Liang}},
  \bibinfo {author} {\bibfnamefont {S.}~\bibnamefont {Ishida}}, \bibinfo
  {author} {\bibfnamefont {Y.}~\bibnamefont {Tomioka}}, \bibinfo {author}
  {\bibfnamefont {K.}~\bibnamefont {Kihou}}, \bibinfo {author} {\bibfnamefont
  {C.~H.}\ \bibnamefont {Lee}}, \bibinfo {author} {\bibfnamefont
  {A.}~\bibnamefont {Iyo}}, \bibinfo {author} {\bibfnamefont {H.}~\bibnamefont
  {Eisaki}}, \bibinfo {author} {\bibfnamefont {T.}~\bibnamefont {Kakeshita}},
  \bibinfo {author} {\bibfnamefont {T.}~\bibnamefont {Ito}}, \ and\ \bibinfo
  {author} {\bibfnamefont {S.}~\bibnamefont {Uchida}},\ }\bibfield  {title}
  {\enquote {\bibinfo {title} {{Unprecedented Anisotropic Metallic State in
  Undoped Iron Arsenide BaFe2As2 Revealed by Optical Spectroscopy}},}\ }\href
  {\doibase 10.1073/pnas.1100102108} {\bibfield  {journal} {\bibinfo  {journal}
  {Proc. Nat. Acad. Sci. U. S. A.}\ }\textbf {\bibinfo {volume} {108}},\
  \bibinfo {pages} {12238} (\bibinfo {year} {2011})}\BibitemShut {NoStop}%
\bibitem [{\citenamefont {Chen}\ \emph {et~al.}(2017)\citenamefont {Chen},
  \citenamefont {Wang}, \citenamefont {Song}, \citenamefont {Lu}, \citenamefont
  {Luo}, \citenamefont {Zhang}, \citenamefont {Dai}, \citenamefont {Yin},
  \citenamefont {Haule},\ and\ \citenamefont {Kotliar}}]{Chen17PRL}%
  \BibitemOpen
  \bibfield  {author} {\bibinfo {author} {\bibfnamefont {Z.-G.}\ \bibnamefont
  {Chen}}, \bibinfo {author} {\bibfnamefont {L.}~\bibnamefont {Wang}}, \bibinfo
  {author} {\bibfnamefont {Y.}~\bibnamefont {Song}}, \bibinfo {author}
  {\bibfnamefont {X.}~\bibnamefont {Lu}}, \bibinfo {author} {\bibfnamefont
  {H.}~\bibnamefont {Luo}}, \bibinfo {author} {\bibfnamefont {C.}~\bibnamefont
  {Zhang}}, \bibinfo {author} {\bibfnamefont {P.}~\bibnamefont {Dai}}, \bibinfo
  {author} {\bibfnamefont {Z.}~\bibnamefont {Yin}}, \bibinfo {author}
  {\bibfnamefont {K.}~\bibnamefont {Haule}}, \ and\ \bibinfo {author}
  {\bibfnamefont {G.}~\bibnamefont {Kotliar}},\ }\bibfield  {title} {\enquote
  {\bibinfo {title} {{Two-Dimensional Massless Dirac Fermions in
  Antiferromagnetic $A{\mathrm{Fe}}_{2}{\mathrm{As}}_{2}$
  $(A=\mathrm{Ba},\mathrm{Sr})$}},}\ }\href {\doibase
  10.1103/PhysRevLett.119.096401} {\bibfield  {journal} {\bibinfo  {journal}
  {Phys. Rev. Lett.}\ }\textbf {\bibinfo {volume} {119}},\ \bibinfo {pages}
  {096401} (\bibinfo {year} {2017})}\BibitemShut {NoStop}%
\bibitem [{\citenamefont {Sebastian}\ \emph {et~al.}(2008)\citenamefont
  {Sebastian}, \citenamefont {Gillett}, \citenamefont {Harrison}, \citenamefont
  {Lau}, \citenamefont {Singh}, \citenamefont {Mielke},\ and\ \citenamefont
  {Lonzarich}}]{Sebastian08JPCM}%
  \BibitemOpen
  \bibfield  {author} {\bibinfo {author} {\bibfnamefont {S.~E.}\ \bibnamefont
  {Sebastian}}, \bibinfo {author} {\bibfnamefont {J.}~\bibnamefont {Gillett}},
  \bibinfo {author} {\bibfnamefont {N.}~\bibnamefont {Harrison}}, \bibinfo
  {author} {\bibfnamefont {P.~H.~C.}\ \bibnamefont {Lau}}, \bibinfo {author}
  {\bibfnamefont {D.~J.}\ \bibnamefont {Singh}}, \bibinfo {author}
  {\bibfnamefont {C.~H.}\ \bibnamefont {Mielke}}, \ and\ \bibinfo {author}
  {\bibfnamefont {G.~G.}\ \bibnamefont {Lonzarich}},\ }\bibfield  {title}
  {\enquote {\bibinfo {title} {{Quantum Oscillations in the Parent Magnetic
  Phase of an Iron Arsenide High Temperature Superconductor}},}\ }\href@noop {}
  {\bibfield  {journal} {\bibinfo  {journal} {J. Phys.: Condens. Matter}\
  }\textbf {\bibinfo {volume} {20}},\ \bibinfo {pages} {422203} (\bibinfo
  {year} {2008})}\BibitemShut {NoStop}%
\bibitem [{\citenamefont {Harrison}\ \emph {et~al.}(2009)\citenamefont
  {Harrison}, \citenamefont {McDonald}, \citenamefont {Mielke}, \citenamefont
  {Bauer}, \citenamefont {Ronning},\ and\ \citenamefont
  {Thompson}}]{Harrison09JPCM}%
  \BibitemOpen
  \bibfield  {author} {\bibinfo {author} {\bibfnamefont {N.}~\bibnamefont
  {Harrison}}, \bibinfo {author} {\bibfnamefont {R.~D.}\ \bibnamefont
  {McDonald}}, \bibinfo {author} {\bibfnamefont {C.~H.}\ \bibnamefont
  {Mielke}}, \bibinfo {author} {\bibfnamefont {E.~D.}\ \bibnamefont {Bauer}},
  \bibinfo {author} {\bibfnamefont {F.}~\bibnamefont {Ronning}}, \ and\
  \bibinfo {author} {\bibfnamefont {J.~D.}\ \bibnamefont {Thompson}},\
  }\bibfield  {title} {\enquote {\bibinfo {title} {{Quantum Oscillations in
  Antiferromagnetic CaFe$_2$As$_2$ on the Brink of Superconductivity}},}\
  }\href@noop {} {\bibfield  {journal} {\bibinfo  {journal} {J. Phys.: Condens.
  Matter}\ }\textbf {\bibinfo {volume} {21}},\ \bibinfo {pages} {322202}
  (\bibinfo {year} {2009})}\BibitemShut {NoStop}%
\bibitem [{\citenamefont {Analytis}\ \emph {et~al.}(2009)\citenamefont
  {Analytis}, \citenamefont {McDonald}, \citenamefont {Chu}, \citenamefont
  {Riggs}, \citenamefont {Bangura}, \citenamefont {Kucharczyk}, \citenamefont
  {Johannes},\ and\ \citenamefont {Fisher}}]{Analytis09PRB}%
  \BibitemOpen
  \bibfield  {author} {\bibinfo {author} {\bibfnamefont {J.~G.}\ \bibnamefont
  {Analytis}}, \bibinfo {author} {\bibfnamefont {R.~D.}\ \bibnamefont
  {McDonald}}, \bibinfo {author} {\bibfnamefont {J.-H.}\ \bibnamefont {Chu}},
  \bibinfo {author} {\bibfnamefont {S.~C.}\ \bibnamefont {Riggs}}, \bibinfo
  {author} {\bibfnamefont {A.~F.}\ \bibnamefont {Bangura}}, \bibinfo {author}
  {\bibfnamefont {C.}~\bibnamefont {Kucharczyk}}, \bibinfo {author}
  {\bibfnamefont {M.}~\bibnamefont {Johannes}}, \ and\ \bibinfo {author}
  {\bibfnamefont {I.~R.}\ \bibnamefont {Fisher}},\ }\bibfield  {title}
  {\enquote {\bibinfo {title} {{Quantum Oscillations in the Parent Pnictide
  BaFe$_2$As$_2$: Itinerant Electrons in the Reconstructed State}},}\ }\href
  {\doibase 10.1103/PhysRevB.80.064507} {\bibfield  {journal} {\bibinfo
  {journal} {Phys. Rev. B}\ }\textbf {\bibinfo {volume} {80}},\ \bibinfo {eid}
  {064507} (\bibinfo {year} {2009})}\BibitemShut {NoStop}%
\bibitem [{\citenamefont {Sutherland}\ \emph {et~al.}(2011)\citenamefont
  {Sutherland}, \citenamefont {Hills}, \citenamefont {Tan}, \citenamefont
  {Altarawneh}, \citenamefont {Harrison}, \citenamefont {Gillett},
  \citenamefont {O'Farrell}, \citenamefont {Benseman}, \citenamefont
  {Kokanovic}, \citenamefont {Syers}, \citenamefont {Cooper},\ and\
  \citenamefont {Sebastian}}]{Sutherland11PRB}%
  \BibitemOpen
  \bibfield  {author} {\bibinfo {author} {\bibfnamefont {M.}~\bibnamefont
  {Sutherland}}, \bibinfo {author} {\bibfnamefont {D.~J.}\ \bibnamefont
  {Hills}}, \bibinfo {author} {\bibfnamefont {B.~S.}\ \bibnamefont {Tan}},
  \bibinfo {author} {\bibfnamefont {M.~M.}\ \bibnamefont {Altarawneh}},
  \bibinfo {author} {\bibfnamefont {N.}~\bibnamefont {Harrison}}, \bibinfo
  {author} {\bibfnamefont {J.}~\bibnamefont {Gillett}}, \bibinfo {author}
  {\bibfnamefont {E.~C.~T.}\ \bibnamefont {O'Farrell}}, \bibinfo {author}
  {\bibfnamefont {T.~M.}\ \bibnamefont {Benseman}}, \bibinfo {author}
  {\bibfnamefont {I.}~\bibnamefont {Kokanovic}}, \bibinfo {author}
  {\bibfnamefont {P.}~\bibnamefont {Syers}}, \bibinfo {author} {\bibfnamefont
  {J.~R.}\ \bibnamefont {Cooper}}, \ and\ \bibinfo {author} {\bibfnamefont
  {S.~E.}\ \bibnamefont {Sebastian}},\ }\bibfield  {title} {\enquote {\bibinfo
  {title} {{Evidence for Dirac Nodes from Quantum Oscillations in
  SrFe${}_{2}$As${}_{2}$}},}\ }\href {\doibase 10.1103/PhysRevB.84.180506}
  {\bibfield  {journal} {\bibinfo  {journal} {Phys. Rev. B}\ }\textbf {\bibinfo
  {volume} {84}},\ \bibinfo {pages} {180506} (\bibinfo {year}
  {2011})}\BibitemShut {NoStop}%
\bibitem [{\citenamefont {Graf}\ \emph {et~al.}(2012)\citenamefont {Graf},
  \citenamefont {Stillwell}, \citenamefont {Murphy}, \citenamefont {Park},
  \citenamefont {Palm}, \citenamefont {Schlottmann}, \citenamefont {McDonald},
  \citenamefont {Analytis}, \citenamefont {Fisher},\ and\ \citenamefont
  {Tozer}}]{Graf12PRB}%
  \BibitemOpen
  \bibfield  {author} {\bibinfo {author} {\bibfnamefont {D.}~\bibnamefont
  {Graf}}, \bibinfo {author} {\bibfnamefont {R.}~\bibnamefont {Stillwell}},
  \bibinfo {author} {\bibfnamefont {T.~P.}\ \bibnamefont {Murphy}}, \bibinfo
  {author} {\bibfnamefont {J.-H.}\ \bibnamefont {Park}}, \bibinfo {author}
  {\bibfnamefont {E.~C.}\ \bibnamefont {Palm}}, \bibinfo {author}
  {\bibfnamefont {P.}~\bibnamefont {Schlottmann}}, \bibinfo {author}
  {\bibfnamefont {R.~D.}\ \bibnamefont {McDonald}}, \bibinfo {author}
  {\bibfnamefont {J.~G.}\ \bibnamefont {Analytis}}, \bibinfo {author}
  {\bibfnamefont {I.~R.}\ \bibnamefont {Fisher}}, \ and\ \bibinfo {author}
  {\bibfnamefont {S.~W.}\ \bibnamefont {Tozer}},\ }\bibfield  {title} {\enquote
  {\bibinfo {title} {{Pressure Dependence of the BaFe${}_{2}$As${}_{2}$ Fermi
  Surface within the Spin Density Wave State}},}\ }\href {\doibase
  10.1103/PhysRevB.85.134503} {\bibfield  {journal} {\bibinfo  {journal} {Phys.
  Rev. B}\ }\textbf {\bibinfo {volume} {85}},\ \bibinfo {pages} {134503}
  (\bibinfo {year} {2012})}\BibitemShut {NoStop}%
\bibitem [{\citenamefont {Rosa}\ \emph {et~al.}(2014)\citenamefont {Rosa},
  \citenamefont {Zeng}, \citenamefont {Adriano}, \citenamefont {Garitezi},
  \citenamefont {Grant}, \citenamefont {Fisk}, \citenamefont {Balicas},
  \citenamefont {Johannes}, \citenamefont {Urbano},\ and\ \citenamefont
  {Pagliuso}}]{Rosa14PRB}%
  \BibitemOpen
  \bibfield  {author} {\bibinfo {author} {\bibfnamefont {P.~F.~S.}\
  \bibnamefont {Rosa}}, \bibinfo {author} {\bibfnamefont {B.}~\bibnamefont
  {Zeng}}, \bibinfo {author} {\bibfnamefont {C.}~\bibnamefont {Adriano}},
  \bibinfo {author} {\bibfnamefont {T.~M.}\ \bibnamefont {Garitezi}}, \bibinfo
  {author} {\bibfnamefont {T.}~\bibnamefont {Grant}}, \bibinfo {author}
  {\bibfnamefont {Z.}~\bibnamefont {Fisk}}, \bibinfo {author} {\bibfnamefont
  {L.}~\bibnamefont {Balicas}}, \bibinfo {author} {\bibfnamefont {M.~D.}\
  \bibnamefont {Johannes}}, \bibinfo {author} {\bibfnamefont {R.~R.}\
  \bibnamefont {Urbano}}, \ and\ \bibinfo {author} {\bibfnamefont {P.~G.}\
  \bibnamefont {Pagliuso}},\ }\bibfield  {title} {\enquote {\bibinfo {title}
  {{Quantum Oscillations in ${\mathrm{EuFe}}_{2}{\mathrm{As}}_{2}$ Single
  Crystals}},}\ }\href {\doibase 10.1103/PhysRevB.90.195146} {\bibfield
  {journal} {\bibinfo  {journal} {Phys. Rev. B}\ }\textbf {\bibinfo {volume}
  {90}},\ \bibinfo {pages} {195146} (\bibinfo {year} {2014})}\BibitemShut
  {NoStop}%
\bibitem [{\citenamefont {Caglieris}\ \emph {et~al.}(2017)\citenamefont
  {Caglieris}, \citenamefont {Leveratto}, \citenamefont {Pallecchi},
  \citenamefont {Bernardini}, \citenamefont {Fujioka}, \citenamefont {Takano},
  \citenamefont {Repetto}, \citenamefont {Jost}, \citenamefont {Zeitler},\ and\
  \citenamefont {Putti}}]{Caglieris17PRB}%
  \BibitemOpen
  \bibfield  {author} {\bibinfo {author} {\bibfnamefont {F.}~\bibnamefont
  {Caglieris}}, \bibinfo {author} {\bibfnamefont {A.}~\bibnamefont
  {Leveratto}}, \bibinfo {author} {\bibfnamefont {I.}~\bibnamefont
  {Pallecchi}}, \bibinfo {author} {\bibfnamefont {F.}~\bibnamefont
  {Bernardini}}, \bibinfo {author} {\bibfnamefont {M.}~\bibnamefont {Fujioka}},
  \bibinfo {author} {\bibfnamefont {Y.}~\bibnamefont {Takano}}, \bibinfo
  {author} {\bibfnamefont {L.}~\bibnamefont {Repetto}}, \bibinfo {author}
  {\bibfnamefont {A.}~\bibnamefont {Jost}}, \bibinfo {author} {\bibfnamefont
  {U.}~\bibnamefont {Zeitler}}, \ and\ \bibinfo {author} {\bibfnamefont
  {M.}~\bibnamefont {Putti}},\ }\bibfield  {title} {\enquote {\bibinfo {title}
  {{Quantum Oscillations in the SmFeAsO Parent Compound and Superconducting
  SmFeAs(O,F)}},}\ }\href {\doibase 10.1103/PhysRevB.96.104508} {\bibfield
  {journal} {\bibinfo  {journal} {Phys. Rev. B}\ }\textbf {\bibinfo {volume}
  {96}},\ \bibinfo {pages} {104508} (\bibinfo {year} {2017})}\BibitemShut
  {NoStop}%
\bibitem [{\citenamefont {Kapitza}(1928)}]{Kapitza28rspa}%
  \BibitemOpen
  \bibfield  {author} {\bibinfo {author} {\bibfnamefont {P.}~\bibnamefont
  {Kapitza}},\ }\bibfield  {title} {\enquote {\bibinfo {title} {{The Study of
  the Specific Resistance of Bismuth Crystals and Its Change in Strong Magnetic
  Fields and Some Allied Problems}},}\ }\href {\doibase 10.1098/rspa.1928.0103}
  {\bibfield  {journal} {\bibinfo  {journal} {Proc. R. Soc. A}\ }\textbf
  {\bibinfo {volume} {119}},\ \bibinfo {pages} {358} (\bibinfo {year}
  {1928})}\BibitemShut {NoStop}%
\bibitem [{\citenamefont {Fuseya}\ \emph {et~al.}(2015)\citenamefont {Fuseya},
  \citenamefont {Ogata},\ and\ \citenamefont {Fukuyama}}]{Fuseya15JPSJ}%
  \BibitemOpen
  \bibfield  {author} {\bibinfo {author} {\bibfnamefont {Y.}~\bibnamefont
  {Fuseya}}, \bibinfo {author} {\bibfnamefont {M.}~\bibnamefont {Ogata}}, \
  and\ \bibinfo {author} {\bibfnamefont {H.}~\bibnamefont {Fukuyama}},\
  }\bibfield  {title} {\enquote {\bibinfo {title} {Transport properties and
  diamagnetism of dirac electrons in bismuth},}\ }\href {\doibase
  10.7566/JPSJ.84.012001} {\bibfield  {journal} {\bibinfo  {journal} {Journal
  of the Physical Society of Japan}\ }\textbf {\bibinfo {volume} {84}},\
  \bibinfo {pages} {012001} (\bibinfo {year} {2015})}\BibitemShut {NoStop}%
\bibitem [{\citenamefont {Liang}\ \emph {et~al.}(2015)\citenamefont {Liang},
  \citenamefont {Gibson}, \citenamefont {Ali}, \citenamefont {Liu},
  \citenamefont {Cava},\ and\ \citenamefont {Ong}}]{Liang15nmat}%
  \BibitemOpen
  \bibfield  {author} {\bibinfo {author} {\bibfnamefont {T.}~\bibnamefont
  {Liang}}, \bibinfo {author} {\bibfnamefont {Q.}~\bibnamefont {Gibson}},
  \bibinfo {author} {\bibfnamefont {M.~N.}\ \bibnamefont {Ali}}, \bibinfo
  {author} {\bibfnamefont {M.}~\bibnamefont {Liu}}, \bibinfo {author}
  {\bibfnamefont {R.~J.}\ \bibnamefont {Cava}}, \ and\ \bibinfo {author}
  {\bibfnamefont {N.~P.}\ \bibnamefont {Ong}},\ }\bibfield  {title} {\enquote
  {\bibinfo {title} {{Ultrahigh Mobility and Giant Magnetoresistance in the
  Dirac Semimetal Cd$_3$As$_2$}},}\ }\href {http://dx.doi.org/10.1038/nmat4143}
  {\bibfield  {journal} {\bibinfo  {journal} {Nat. Mater.}\ }\textbf {\bibinfo
  {volume} {14}},\ \bibinfo {pages} {280} (\bibinfo {year} {2015})}\BibitemShut
  {NoStop}%
\bibitem [{\citenamefont {He}\ \emph {et~al.}(2014)\citenamefont {He},
  \citenamefont {Hong}, \citenamefont {Dong}, \citenamefont {Pan},
  \citenamefont {Zhang}, \citenamefont {Zhang},\ and\ \citenamefont
  {Li}}]{He14PRL}%
  \BibitemOpen
  \bibfield  {author} {\bibinfo {author} {\bibfnamefont {L.~P.}\ \bibnamefont
  {He}}, \bibinfo {author} {\bibfnamefont {X.~C.}\ \bibnamefont {Hong}},
  \bibinfo {author} {\bibfnamefont {J.~K.}\ \bibnamefont {Dong}}, \bibinfo
  {author} {\bibfnamefont {J.}~\bibnamefont {Pan}}, \bibinfo {author}
  {\bibfnamefont {Z.}~\bibnamefont {Zhang}}, \bibinfo {author} {\bibfnamefont
  {J.}~\bibnamefont {Zhang}}, \ and\ \bibinfo {author} {\bibfnamefont {S.~Y.}\
  \bibnamefont {Li}},\ }\bibfield  {title} {\enquote {\bibinfo {title}
  {{Quantum Transport Evidence for the Three-Dimensional Dirac Semimetal Phase
  in ${\mathrm{Cd}}_{3}{\mathrm{As}}_{2}$}},}\ }\href {\doibase
  10.1103/PhysRevLett.113.246402} {\bibfield  {journal} {\bibinfo  {journal}
  {Phys. Rev. Lett.}\ }\textbf {\bibinfo {volume} {113}},\ \bibinfo {pages}
  {246402} (\bibinfo {year} {2014})}\BibitemShut {NoStop}%
\bibitem [{\citenamefont {Feng}\ \emph {et~al.}(2015)\citenamefont {Feng},
  \citenamefont {Pang}, \citenamefont {Wu}, \citenamefont {Wang}, \citenamefont
  {Weng}, \citenamefont {Li}, \citenamefont {Dai}, \citenamefont {Fang},
  \citenamefont {Shi},\ and\ \citenamefont {Lu}}]{Feng15PRB}%
  \BibitemOpen
  \bibfield  {author} {\bibinfo {author} {\bibfnamefont {J.}~\bibnamefont
  {Feng}}, \bibinfo {author} {\bibfnamefont {Y.}~\bibnamefont {Pang}}, \bibinfo
  {author} {\bibfnamefont {D.}~\bibnamefont {Wu}}, \bibinfo {author}
  {\bibfnamefont {Z.}~\bibnamefont {Wang}}, \bibinfo {author} {\bibfnamefont
  {H.}~\bibnamefont {Weng}}, \bibinfo {author} {\bibfnamefont {J.}~\bibnamefont
  {Li}}, \bibinfo {author} {\bibfnamefont {X.}~\bibnamefont {Dai}}, \bibinfo
  {author} {\bibfnamefont {Z.}~\bibnamefont {Fang}}, \bibinfo {author}
  {\bibfnamefont {Y.}~\bibnamefont {Shi}}, \ and\ \bibinfo {author}
  {\bibfnamefont {L.}~\bibnamefont {Lu}},\ }\bibfield  {title} {\enquote
  {\bibinfo {title} {{Large Linear Magnetoresistance in Dirac Semimetal
  ${\mathrm{Cd}}_{3}{\mathrm{As}}_{2}$ with Fermi Surfaces Close to the Dirac
  Points}},}\ }\href {\doibase 10.1103/PhysRevB.92.081306} {\bibfield
  {journal} {\bibinfo  {journal} {Phys. Rev. B}\ }\textbf {\bibinfo {volume}
  {92}},\ \bibinfo {pages} {081306} (\bibinfo {year} {2015})}\BibitemShut
  {NoStop}%
\bibitem [{\citenamefont {Kushwaha}\ \emph {et~al.}(2015)\citenamefont
  {Kushwaha}, \citenamefont {Krizan}, \citenamefont {Feldman}, \citenamefont
  {Gyenis}, \citenamefont {Randeria}, \citenamefont {Xiong}, \citenamefont
  {Xu}, \citenamefont {Alidoust}, \citenamefont {Belopolski}, \citenamefont
  {Liang}, \citenamefont {Hasan}, \citenamefont {Ong}, \citenamefont
  {Yazdani},\ and\ \citenamefont {Cava}}]{Kushwaha15APLMaterials}%
  \BibitemOpen
  \bibfield  {author} {\bibinfo {author} {\bibfnamefont {S.~K.}\ \bibnamefont
  {Kushwaha}}, \bibinfo {author} {\bibfnamefont {J.~W.}\ \bibnamefont
  {Krizan}}, \bibinfo {author} {\bibfnamefont {B.~E.}\ \bibnamefont {Feldman}},
  \bibinfo {author} {\bibfnamefont {A.}~\bibnamefont {Gyenis}}, \bibinfo
  {author} {\bibfnamefont {M.~T.}\ \bibnamefont {Randeria}}, \bibinfo {author}
  {\bibfnamefont {J.}~\bibnamefont {Xiong}}, \bibinfo {author} {\bibfnamefont
  {S.-Y.}\ \bibnamefont {Xu}}, \bibinfo {author} {\bibfnamefont
  {N.}~\bibnamefont {Alidoust}}, \bibinfo {author} {\bibfnamefont
  {I.}~\bibnamefont {Belopolski}}, \bibinfo {author} {\bibfnamefont
  {T.}~\bibnamefont {Liang}}, \bibinfo {author} {\bibfnamefont {M.~Z.}\
  \bibnamefont {Hasan}}, \bibinfo {author} {\bibfnamefont {N.~P.}\ \bibnamefont
  {Ong}}, \bibinfo {author} {\bibfnamefont {A.}~\bibnamefont {Yazdani}}, \ and\
  \bibinfo {author} {\bibfnamefont {R.~J.}\ \bibnamefont {Cava}},\ }\bibfield
  {title} {\enquote {\bibinfo {title} {{Bulk Crystal Growth and Electronic
  Characterization of the 3D Dirac Semimetal Na$_3$Bi}},}\ }\href {\doibase
  10.1063/1.4908158} {\bibfield  {journal} {\bibinfo  {journal} {APL
  Materials}\ }\textbf {\bibinfo {volume} {3}},\ \bibinfo {pages} {041504}
  (\bibinfo {year} {2015})}\BibitemShut {NoStop}%
\bibitem [{\citenamefont {Novak}\ \emph {et~al.}(2015)\citenamefont {Novak},
  \citenamefont {Sasaki}, \citenamefont {Segawa},\ and\ \citenamefont
  {Ando}}]{Novak15PRB}%
  \BibitemOpen
  \bibfield  {author} {\bibinfo {author} {\bibfnamefont {M.}~\bibnamefont
  {Novak}}, \bibinfo {author} {\bibfnamefont {S.}~\bibnamefont {Sasaki}},
  \bibinfo {author} {\bibfnamefont {K.}~\bibnamefont {Segawa}}, \ and\ \bibinfo
  {author} {\bibfnamefont {Y.}~\bibnamefont {Ando}},\ }\bibfield  {title}
  {\enquote {\bibinfo {title} {{Large Linear Magnetoresistance in the Dirac
  Semimetal TlBiSSe}},}\ }\href {\doibase 10.1103/PhysRevB.91.041203}
  {\bibfield  {journal} {\bibinfo  {journal} {Phys. Rev. B}\ }\textbf {\bibinfo
  {volume} {91}},\ \bibinfo {pages} {041203} (\bibinfo {year}
  {2015})}\BibitemShut {NoStop}%
\bibitem [{\citenamefont {Singha}\ \emph {et~al.}(2017)\citenamefont {Singha},
  \citenamefont {Pariari}, \citenamefont {Satpati},\ and\ \citenamefont
  {Mandal}}]{Singha17PNAS}%
  \BibitemOpen
  \bibfield  {author} {\bibinfo {author} {\bibfnamefont {R.}~\bibnamefont
  {Singha}}, \bibinfo {author} {\bibfnamefont {A.~K.}\ \bibnamefont {Pariari}},
  \bibinfo {author} {\bibfnamefont {B.}~\bibnamefont {Satpati}}, \ and\
  \bibinfo {author} {\bibfnamefont {P.}~\bibnamefont {Mandal}},\ }\bibfield
  {title} {\enquote {\bibinfo {title} {{Large Nonsaturating Magnetoresistance
  and Signature of Nondegenerate Dirac Nodes in ZrSiS}},}\ }\href {\doibase
  10.1073/pnas.1618004114} {\bibfield  {journal} {\bibinfo  {journal} {Proc.
  Nat. Acad. Sci. U. S. A.}\ }\textbf {\bibinfo {volume} {114}},\ \bibinfo
  {pages} {2468} (\bibinfo {year} {2017})}\BibitemShut {NoStop}%
\bibitem [{\citenamefont {{Pallecchi, I.}}\ \emph {et~al.}(2013)\citenamefont
  {{Pallecchi, I.}}, \citenamefont {{Bernardini, F.}}, \citenamefont
  {{Caglieris, F.}}, \citenamefont {{Palenzona, A.}}, \citenamefont {{Massidda,
  S.}},\ and\ \citenamefont {{Putti, M.}}}]{Pallecchi13EPJB}%
  \BibitemOpen
  \bibfield  {author} {\bibinfo {author} {\bibnamefont {{Pallecchi, I.}}},
  \bibinfo {author} {\bibnamefont {{Bernardini, F.}}}, \bibinfo {author}
  {\bibnamefont {{Caglieris, F.}}}, \bibinfo {author} {\bibnamefont
  {{Palenzona, A.}}}, \bibinfo {author} {\bibnamefont {{Massidda, S.}}}, \ and\
  \bibinfo {author} {\bibnamefont {{Putti, M.}}},\ }\bibfield  {title}
  {\enquote {\bibinfo {title} {{Role of Dirac Cones in Magnetotransport
  Properties of REFeAsO (RE = rare earth) Oxypnictides}},}\ }\href {\doibase
  10.1140/epjb/e2013-40148-6} {\bibfield  {journal} {\bibinfo  {journal} {Eur.
  Phys. J. B}\ }\textbf {\bibinfo {volume} {86}},\ \bibinfo {pages} {338}
  (\bibinfo {year} {2013})}\BibitemShut {NoStop}%
\bibitem [{\citenamefont {Abrikosov}(1998)}]{Abrikosov98PRB}%
  \BibitemOpen
  \bibfield  {author} {\bibinfo {author} {\bibfnamefont {A.~A.}\ \bibnamefont
  {Abrikosov}},\ }\bibfield  {title} {\enquote {\bibinfo {title} {{Quantum
  Magnetoresistance}},}\ }\href {\doibase 10.1103/PhysRevB.58.2788} {\bibfield
  {journal} {\bibinfo  {journal} {Phys. Rev. B}\ }\textbf {\bibinfo {volume}
  {58}},\ \bibinfo {pages} {2788} (\bibinfo {year} {1998})}\BibitemShut
  {NoStop}%
\bibitem [{\citenamefont {Vildosola}\ \emph {et~al.}(2008)\citenamefont
  {Vildosola}, \citenamefont {Pourovskii}, \citenamefont {Arita}, \citenamefont
  {Biermann},\ and\ \citenamefont {Georges}}]{Vildosola08PRB}%
  \BibitemOpen
  \bibfield  {author} {\bibinfo {author} {\bibfnamefont {V.}~\bibnamefont
  {Vildosola}}, \bibinfo {author} {\bibfnamefont {L.}~\bibnamefont
  {Pourovskii}}, \bibinfo {author} {\bibfnamefont {R.}~\bibnamefont {Arita}},
  \bibinfo {author} {\bibfnamefont {S.}~\bibnamefont {Biermann}}, \ and\
  \bibinfo {author} {\bibfnamefont {A.}~\bibnamefont {Georges}},\ }\bibfield
  {title} {\enquote {\bibinfo {title} {{Bandwidth and Fermi Surface of Iron
  Oxypnictides: Covalency and Sensitivity to Structural Changes}},}\ }\href
  {\doibase 10.1103/PhysRevB.78.064518} {\bibfield  {journal} {\bibinfo
  {journal} {Phys. Rev. B}\ }\textbf {\bibinfo {volume} {78}},\ \bibinfo
  {pages} {064518} (\bibinfo {year} {2008})}\BibitemShut {NoStop}%
\bibitem [{\citenamefont {Kuroki}\ \emph {et~al.}(2009)\citenamefont {Kuroki},
  \citenamefont {Usui}, \citenamefont {Onari}, \citenamefont {Arita},\ and\
  \citenamefont {Aoki}}]{Kuroki09PRB}%
  \BibitemOpen
  \bibfield  {author} {\bibinfo {author} {\bibfnamefont {K.}~\bibnamefont
  {Kuroki}}, \bibinfo {author} {\bibfnamefont {H.}~\bibnamefont {Usui}},
  \bibinfo {author} {\bibfnamefont {S.}~\bibnamefont {Onari}}, \bibinfo
  {author} {\bibfnamefont {R.}~\bibnamefont {Arita}}, \ and\ \bibinfo {author}
  {\bibfnamefont {H.}~\bibnamefont {Aoki}},\ }\bibfield  {title} {\enquote
  {\bibinfo {title} {{Pnictogen Height as a Possible Switch Between
  High-${T}_{c}$ Nodeless and Low-${T}_{c}$ Nodal Pairings in the Iron-Based
  Superconductors}},}\ }\href {\doibase 10.1103/PhysRevB.79.224511} {\bibfield
  {journal} {\bibinfo  {journal} {Phys. Rev. B}\ }\textbf {\bibinfo {volume}
  {79}},\ \bibinfo {pages} {224511} (\bibinfo {year} {2009})}\BibitemShut
  {NoStop}%
\bibitem [{\citenamefont {de~la Cruz}\ \emph {et~al.}(2008)\citenamefont {de~la
  Cruz}, \citenamefont {Huang}, \citenamefont {Lynn}, \citenamefont {Li},
  \citenamefont {II}, \citenamefont {Zarestky}, \citenamefont {Mook},
  \citenamefont {Chen}, \citenamefont {Luo}, \citenamefont {Wang},\ and\
  \citenamefont {Dai}}]{Cruz08nature}%
  \BibitemOpen
  \bibfield  {author} {\bibinfo {author} {\bibfnamefont {C.}~\bibnamefont
  {de~la Cruz}}, \bibinfo {author} {\bibfnamefont {Q.}~\bibnamefont {Huang}},
  \bibinfo {author} {\bibfnamefont {J.~W.}\ \bibnamefont {Lynn}}, \bibinfo
  {author} {\bibfnamefont {J.}~\bibnamefont {Li}}, \bibinfo {author}
  {\bibfnamefont {W.~R.}\ \bibnamefont {II}}, \bibinfo {author} {\bibfnamefont
  {J.~L.}\ \bibnamefont {Zarestky}}, \bibinfo {author} {\bibfnamefont {H.~A.}\
  \bibnamefont {Mook}}, \bibinfo {author} {\bibfnamefont {G.~F.}\ \bibnamefont
  {Chen}}, \bibinfo {author} {\bibfnamefont {J.~L.}\ \bibnamefont {Luo}},
  \bibinfo {author} {\bibfnamefont {N.~L.}\ \bibnamefont {Wang}}, \ and\
  \bibinfo {author} {\bibfnamefont {P.}~\bibnamefont {Dai}},\ }\bibfield
  {title} {\enquote {\bibinfo {title} {{Magnetic Order Close to
  Superconductivity in the Iron-Based Layered LaO$_{1-x}$F$_x$FeAs Systems}},}\
  }\href {http://dx.doi.org/10.1038/nature07057} {\bibfield  {journal}
  {\bibinfo  {journal} {Nature}\ }\textbf {\bibinfo {volume} {453}},\ \bibinfo
  {pages} {899} (\bibinfo {year} {2008})}\BibitemShut {NoStop}%
\bibitem [{\citenamefont {Ma}\ \emph {et~al.}(2016)\citenamefont {Ma},
  \citenamefont {Hu}, \citenamefont {Ji}, \citenamefont {Gao}, \citenamefont
  {Zhang}, \citenamefont {Mu}, \citenamefont {Huang},\ and\ \citenamefont
  {Xie}}]{MA16JCG}%
  \BibitemOpen
  \bibfield  {author} {\bibinfo {author} {\bibfnamefont {Y.}~\bibnamefont
  {Ma}}, \bibinfo {author} {\bibfnamefont {K.}~\bibnamefont {Hu}}, \bibinfo
  {author} {\bibfnamefont {Q.}~\bibnamefont {Ji}}, \bibinfo {author}
  {\bibfnamefont {B.}~\bibnamefont {Gao}}, \bibinfo {author} {\bibfnamefont
  {H.}~\bibnamefont {Zhang}}, \bibinfo {author} {\bibfnamefont
  {G.}~\bibnamefont {Mu}}, \bibinfo {author} {\bibfnamefont {F.}~\bibnamefont
  {Huang}}, \ and\ \bibinfo {author} {\bibfnamefont {X.}~\bibnamefont {Xie}},\
  }\bibfield  {title} {\enquote {\bibinfo {title} {{Growth and Characterization
  of CaFe$_{1-x}$Co$_x$AsF Single Crystals by CaAs Flux Method}},}\ }\href
  {\doibase https://doi.org/10.1016/j.jcrysgro.2016.07.029} {\bibfield
  {journal} {\bibinfo  {journal} {J. Cryst. Growth}\ }\textbf {\bibinfo
  {volume} {451}},\ \bibinfo {pages} {161 } (\bibinfo {year}
  {2016})}\BibitemShut {NoStop}%
\bibitem [{\citenamefont {Dong}\ \emph {et~al.}(2008)\citenamefont {Dong},
  \citenamefont {Zhang}, \citenamefont {Xu}, \citenamefont {Li}, \citenamefont
  {Li}, \citenamefont {Hu}, \citenamefont {Wu}, \citenamefont {Chen},
  \citenamefont {Dai}, \citenamefont {Luo}, \citenamefont {Fang},\ and\
  \citenamefont {Wang}}]{Dong08EPL}%
  \BibitemOpen
  \bibfield  {author} {\bibinfo {author} {\bibfnamefont {J.}~\bibnamefont
  {Dong}}, \bibinfo {author} {\bibfnamefont {H.~J.}\ \bibnamefont {Zhang}},
  \bibinfo {author} {\bibfnamefont {G.}~\bibnamefont {Xu}}, \bibinfo {author}
  {\bibfnamefont {Z.}~\bibnamefont {Li}}, \bibinfo {author} {\bibfnamefont
  {G.}~\bibnamefont {Li}}, \bibinfo {author} {\bibfnamefont {W.~Z.}\
  \bibnamefont {Hu}}, \bibinfo {author} {\bibfnamefont {D.}~\bibnamefont {Wu}},
  \bibinfo {author} {\bibfnamefont {G.~F.}\ \bibnamefont {Chen}}, \bibinfo
  {author} {\bibfnamefont {X.}~\bibnamefont {Dai}}, \bibinfo {author}
  {\bibfnamefont {J.~L.}\ \bibnamefont {Luo}}, \bibinfo {author} {\bibfnamefont
  {Z.}~\bibnamefont {Fang}}, \ and\ \bibinfo {author} {\bibfnamefont {N.~L.}\
  \bibnamefont {Wang}},\ }\bibfield  {title} {\enquote {\bibinfo {title}
  {{Competing Orders and Spin-Density-Wave Instability in
  La(O$_{1-x}$F$_x$)FeAs}},}\ }\href
  {http://stacks.iop.org/0295-5075/83/i=2/a=27006} {\bibfield  {journal}
  {\bibinfo  {journal} {EPL}\ }\textbf {\bibinfo {volume} {83}},\ \bibinfo
  {pages} {27006} (\bibinfo {year} {2008})}\BibitemShut {NoStop}%
\bibitem [{\citenamefont {Kohama}\ \emph {et~al.}(2008)\citenamefont {Kohama},
  \citenamefont {Kamihara}, \citenamefont {Hirano}, \citenamefont {Kawaji},
  \citenamefont {Atake},\ and\ \citenamefont {Hosono}}]{Kohama08PRB}%
  \BibitemOpen
  \bibfield  {author} {\bibinfo {author} {\bibfnamefont {Y.}~\bibnamefont
  {Kohama}}, \bibinfo {author} {\bibfnamefont {Y.}~\bibnamefont {Kamihara}},
  \bibinfo {author} {\bibfnamefont {M.}~\bibnamefont {Hirano}}, \bibinfo
  {author} {\bibfnamefont {H.}~\bibnamefont {Kawaji}}, \bibinfo {author}
  {\bibfnamefont {T.}~\bibnamefont {Atake}}, \ and\ \bibinfo {author}
  {\bibfnamefont {H.}~\bibnamefont {Hosono}},\ }\bibfield  {title} {\enquote
  {\bibinfo {title} {{Ferromagnetic Spin Fluctuation in
  ${\text{LaFeAsO}}_{1\ensuremath{-}x}{\text{F}}_{x}$}},}\ }\href {\doibase
  10.1103/PhysRevB.78.020512} {\bibfield  {journal} {\bibinfo  {journal} {Phys.
  Rev. B}\ }\textbf {\bibinfo {volume} {78}},\ \bibinfo {pages} {020512}
  (\bibinfo {year} {2008})}\BibitemShut {NoStop}%
\bibitem [{\citenamefont {Zhigadlo}\ \emph {et~al.}(2012)\citenamefont
  {Zhigadlo}, \citenamefont {Weyeneth}, \citenamefont {Katrych}, \citenamefont
  {Moll}, \citenamefont {Rogacki}, \citenamefont {Bosma}, \citenamefont
  {Puzniak}, \citenamefont {Karpinski},\ and\ \citenamefont
  {Batlogg}}]{Zhigadlo12PRB}%
  \BibitemOpen
  \bibfield  {author} {\bibinfo {author} {\bibfnamefont {N.~D.}\ \bibnamefont
  {Zhigadlo}}, \bibinfo {author} {\bibfnamefont {S.}~\bibnamefont {Weyeneth}},
  \bibinfo {author} {\bibfnamefont {S.}~\bibnamefont {Katrych}}, \bibinfo
  {author} {\bibfnamefont {P.~J.~W.}\ \bibnamefont {Moll}}, \bibinfo {author}
  {\bibfnamefont {K.}~\bibnamefont {Rogacki}}, \bibinfo {author} {\bibfnamefont
  {S.}~\bibnamefont {Bosma}}, \bibinfo {author} {\bibfnamefont
  {R.}~\bibnamefont {Puzniak}}, \bibinfo {author} {\bibfnamefont
  {J.}~\bibnamefont {Karpinski}}, \ and\ \bibinfo {author} {\bibfnamefont
  {B.}~\bibnamefont {Batlogg}},\ }\bibfield  {title} {\enquote {\bibinfo
  {title} {{High-Pressure Flux Growth, Structural, and Superconducting
  Properties of $Ln$FeAsO ($Ln$ $=$ Pr, Nd, Sm) Single Crystals}},}\ }\href
  {\doibase 10.1103/PhysRevB.86.214509} {\bibfield  {journal} {\bibinfo
  {journal} {Phys. Rev. B}\ }\textbf {\bibinfo {volume} {86}},\ \bibinfo
  {pages} {214509} (\bibinfo {year} {2012})}\BibitemShut {NoStop}%
\end{thebibliography}
%

\end{document}